 \documentclass[11pt]{article}

 \pdfoutput=1

\usepackage[round]{natbib}

\usepackage[english]{babel}

\usepackage{graphicx} 
\usepackage{amsmath}

\usepackage{enumitem}

\usepackage{float}
\usepackage{ifthen}
\usepackage{hyperref}

\usepackage{titlesec}

\titlespacing*{\section}{0pt}{1.1\baselineskip}{\baselineskip}

\usepackage{color}

\begin{document}

\def\hmath#1{\text{\scalebox{1.5}{$#1$}}}
\def\lmath#1{\text{\scalebox{1.4}{$#1$}}}
\def\mmath#1{\text{\scalebox{1.2}{$#1$}}}
\def\smath#1{\text{\scalebox{.8}{$#1$}}}

\def\hfrac#1#2{\hmath{\frac{#1}{#2}}}
\def\lfrac#1#2{\lmath{\frac{#1}{#2}}}
\def\mfrac#1#2{\mmath{\frac{#1}{#2}}}
\def\sfrac#1#2{\smath{\frac{#1}{#2}}}

\def\pow{^\mmath}

\title{\vspace{-2mm} Comment on the paper ``Generalized dynamic 
equations related to condensation and freezing 
processes'' by Wang and Huang (2018).}

\author{\vspace{-2mm} by Pascal Marquet {\it M\'et\'eo-France}}


\date{\vspace{-4mm} \today}

\maketitle

\vspace{-8mm}
\begin{center}
{\em Paper submitted on the 26th of  September 2018 to the
\underline{Journal of Geophysical Research: Atmospheres}. }
\vspace{2mm}\\
{\em \underline{Corresponding address}: pascal.marquet@meteo.fr}
\end{center}
\vspace{-2mm}


\begin{center}
{Partial English translations of 4 Chinese papers of {\bf Xing-Rong Wang} are provided:}
\vspace{1mm}\\
{in the Appendix~1 for }{\bf Wang and Wu (1995)}
\vspace{1mm}\\
{in the Appendix~2 for }{\bf Wang, Wang and Shi (1998)}
\vspace{1mm}\\
{in the Appendix~3 for }{\bf Wang, Wu and Shi (1999)}
\vspace{1mm}\\
{in the Appendix~4 for }{\bf Wang and Wei (2007)}
\end{center}
\vspace{-4mm}




 \section{The Condensation Probability Function is Arbitrary}
\label{section_CPF}
\vspace{-2mm}


A Condensation Probability Function denoted by ``$r^k$'' 
is defined in 
\cite[hereafter WH18]{Wang_al_2018}
 in accordance with to the previous Chinese papers by \citet[hereafter WW95]{Wang_Wu_95} 
and \citet[hereafter WWS99]{Wang_al_99}.
This function was defined first as $K(q/q_s) = (q/q_s)^{\alpha}$ in Eq.~(13) of WW95 to conform to the constraints: 
i) $K(0)=0$; and
ii) $K(1)=1$.
However, there are many ways for a function $K(q, q_s)$ to conform to these  constraints, and the motivations for the special formulation $(q/q_s)^\alpha$ chosen in WW95 and maintain in WH18 are not clearly exposed.
This is a first issue.

Once the arbitrary function $K=(q/q_s)^k$ had been chosen in WW95, the special value $k = 9$ was tuned from Table~1 in WWS99 to conform to the additional constraints: 
iii) $(q/q_s)^k$ must decrease ``sufficiently rapidly'' for relative humidities $RH = 100 \; (q/q_s)$ lower than ``a certain threshold of $78$~\%''.

\begin{figure}[htb]
\centering
\includegraphics[width=0.95\linewidth]{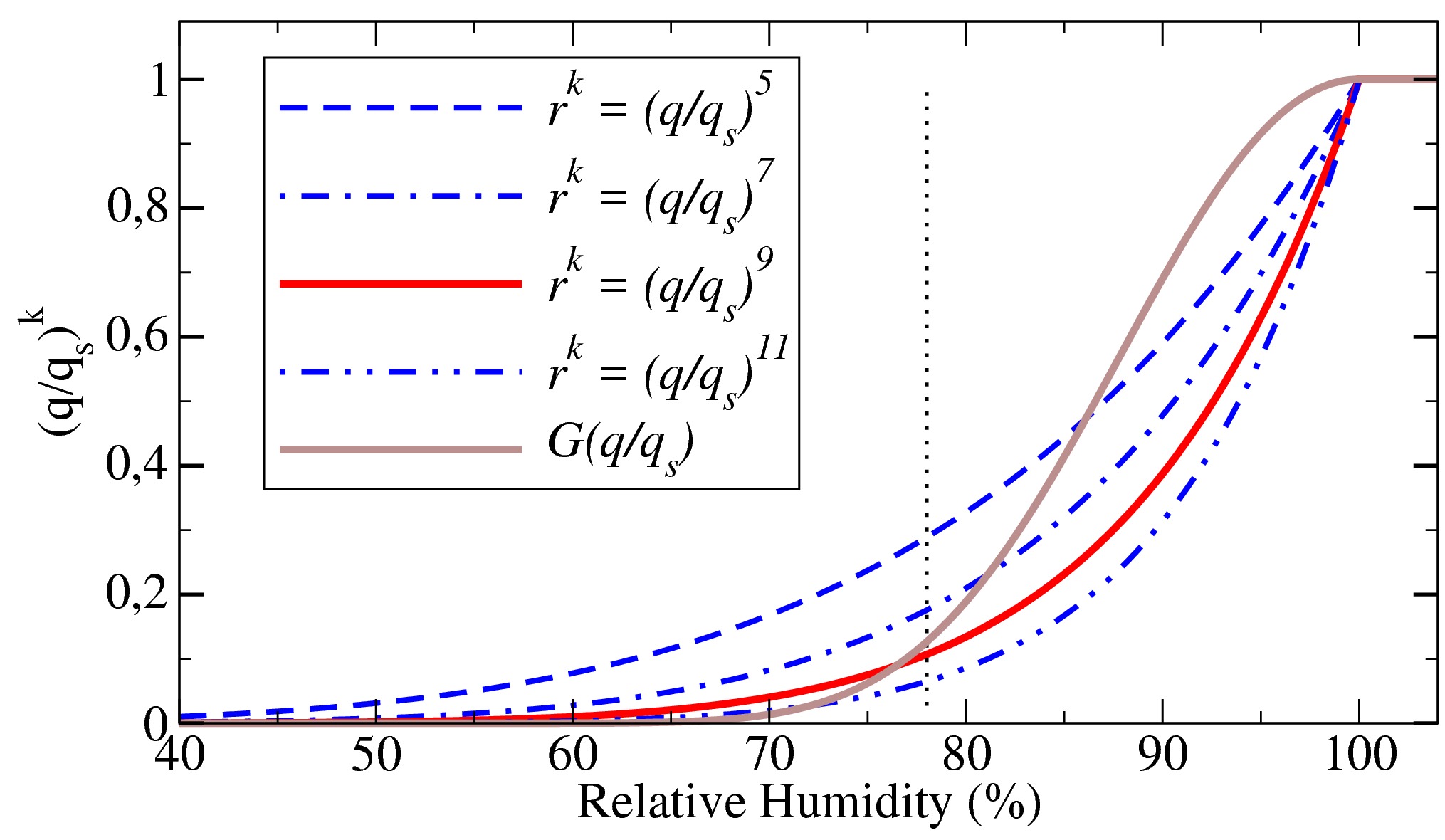}
\vspace{-2mm}
\caption{\small \it
Plot of the ``Condensation Probability Function'' denoted by $r^k = (q/q_s)^k$ in WH18, with the relative humidity $100 \: (q/q_s)$ in abscissa.
The curves for $k=(5,7,11)$ are in blue and the one for $k=9$ is in red.
The greyish curve is for the function $G(q/q_s)$ given by Eq.(\ref{eq_GX}).
}
\vspace{0mm}
\label{fig_HU}
\end{figure}

However, the curves of the function $(q/q_s)^k$, depending on $k$ and $RH$, are not plotted in WW95, WWS99 or WH18.
These curves are plotted in Fig.\ref{fig_HU}, which clearly shows that the requirement iii) is satisfied with any value of $k$ between $7$ and $11$, but without any criteria given in WWS99 to explain why $k=9$ is the most relevant tuning.

A possible, smoother function, $G$, is plotted in Fig.\ref{fig_HU} according to
\begin{equation}
G(X) = 
\exp\left[\:
- \: {(X-1)^2} \: / \: {( \, \beta \: X \, )}
\:\right] \: ,
\label{eq_GX}
\end{equation}
where $X=q/q_s$.
This function has the same properties $G\,'(0) = 0$, $G(0)=0$ and $G(1)=0$ required by WW95 and kept in WH18 with, additionally, $F(X_0) \approx G(X_0)$ for $X_0 = 78$~\% if the tuning parameter is set to $\beta=0.03$.
The additional interesting property is $G\,'(1) = 0$, which means that the null derivative is continuous for $RH = 100$~\%, differently from 
$K\,'(1-\varepsilon) \approx k = 9 \gg 0$ for small $\varepsilon >0$.
This property is used in section~\ref{section_pot_temp} to derive a smoother version of the potential temperature defined in WH18.



According to all the Chinese and English papers by Xingrong Wang, the aim of introducing the function $K = (q/q_s)^k$ appears to ``eliminate the irrational implicit assumption'' that condensation would occur in the atmosphere at the threshold of $100$~\% of relative humidity.
Xingrong Wang explains that he uses the ``observational evidence presented by 
\citet{Mason_1971}''
that the condensation process may occur at lower values, such as $X_0 = 78$~\%.
However, a careful reading of 
\citet[p.28-29]{Mason_1971}
reveals that this threshold of $X_0$ only concerns (aerosol) nuclei of pure NaCl hygroscopic salt particles.
The same opinion is given in
\citet[p.31]{Doswell_1985},
where it is explained that 
only the ``oceanic salt (resulting when sea spray evaporates in the air)'' is hygroscopic and is soluble in water and acts to lower the equilibrium vapor pressure.

Therefore, this threshold of about $78$~\% cannot be universal throughout the atmosphere, that is to say valid everywhere, at all levels, for all temperatures.
The vertical and horizontal distributions of the salt/marine aerosols are highly variable from place to place \citep{Bozzo_al_2017}, making the universal function $K=(q/q_s)^k$ plotted in Fig.\ref{fig_HU} unrealistic, since it is the same everywhere in the atmosphere in all papers by Xingrong Wang, including in \citet{Gao_al_2004}.
In particular, it is likely that the threshold of $100$~\% must be applied to initiate condensation in mid- and high-tropospheric clouds, where seal-salt aerosols have tiny concentrations.


Another issue is the sentence written in the abstract and the conclusion  of the paper by \citet{Wang_Feng_2015}: the motivation for the introduction of the ``Condensation Probability Function'' $(q/q_s)^k$ used in WH18 would be
``to eliminate the irrational supposition that condensed liquid water always falls immediately''.
However, even if it was considered acceptable in the 1980s and 1990s, and maybe up to the mid 2000s, this assumption is no longer made nowadays and the use of prognostic microphysical schemes has become very common, even widespread, in all modern GCM or NWP models.

The authors could have addressed the link between the function $K=(q/q_s)^k$ and the sub-grid variability of water, as used in all modern GCM and NWP models. 
The method described first in 
\citet{Sommeria_Deardorff_1977} 
corresponds to the use of diagnostic (or prognostic) PDF functions to allow the grid cells to generate condensate (liquid water or ice) well before the threshold of $100$~\% of relative humidity.
This method is thus similar to the use of $K=(q/q_s)^k$, but the motivations are different, and they do not correspond to the same physical meaning.

 \section{A Latent Heat Cannot Be Continuous}
\label{section_latent_heat}
\vspace{-2mm}

A new formulation for the latent heat of moist air is suggested in WH18.
It is defined by:
\begin{equation}
L^{\ast}(T) 
\; = \;
L_i \: \phi(T) \: + \: L_d  
\; = \; 
L_{fus}(T) \: \varphi(T) \: + \: L_{vap}(T) \: ,
\label{eq_LWH18}
\end{equation}
where the right hand side uses more usual meteorological notations:
the term $L_i$ is the latent heat of fusion $L_{fus} = L_{sub} - L_{vap}$ and the term $L_d$ is the latent heat of vaporization $L_{vap}$.
The notation $\varphi(T)$ is used in WH18  to represent a freezing probability function defined by a Cumulative Distribution Function, with a tuning parameter $\sigma$ determined by  the hypothesis:
$\varphi(277.13) = 0.05$ at $+3.98\:{}^{\circ}$C$ \, \approx +4\:{}^{\circ}$C.


\begin{figure}[h]
\centering
\includegraphics[width=0.95\linewidth]{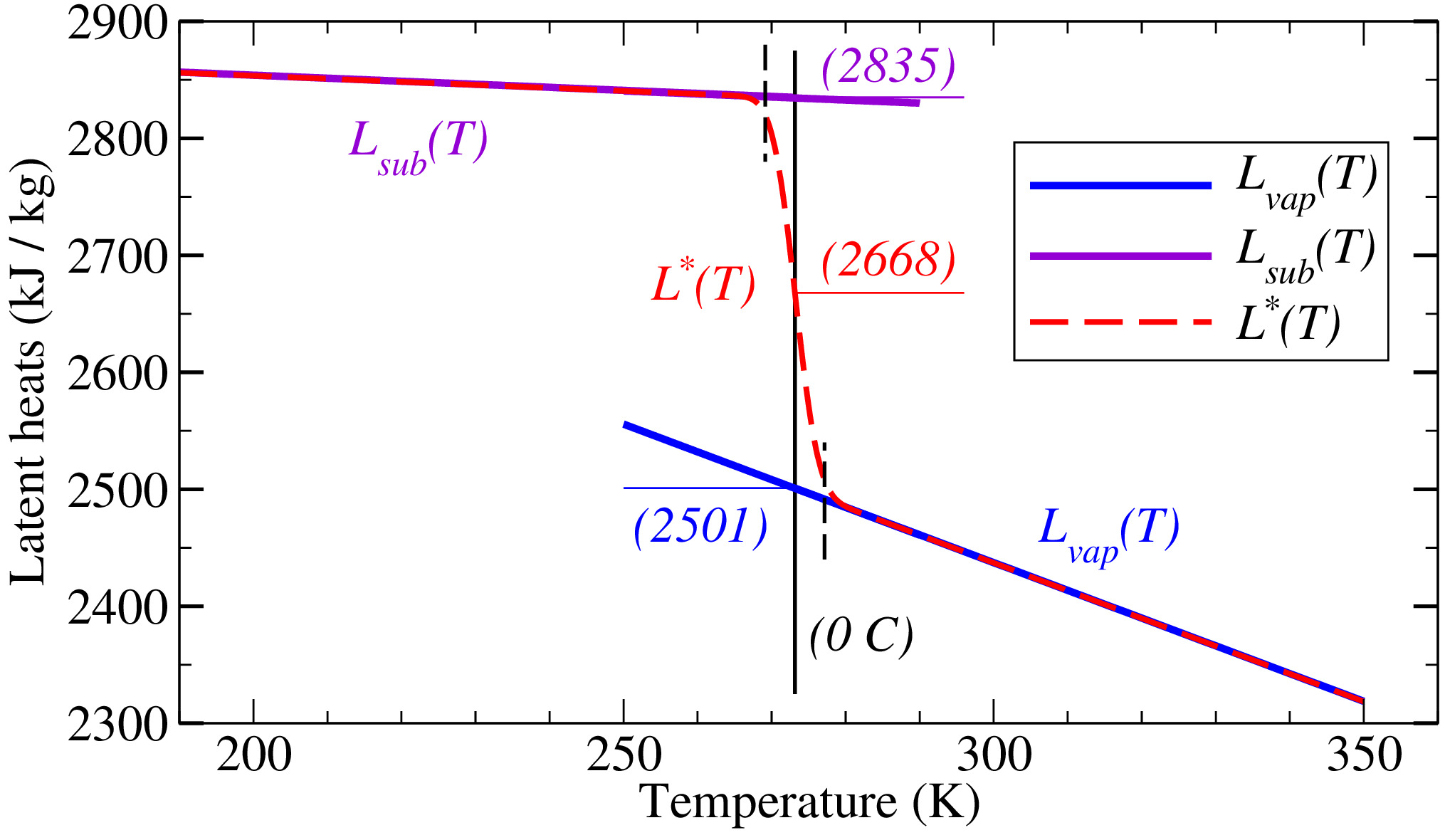}
\vspace{-2mm}
\caption{\small \it
The usual latent heats of vaporization $L_{vap}(T)$ (in blue) and sublimation $L_{sub}(T)$ (in purple). 
The new latent heat $L^{\ast}(T)$ of WH18 is represented by the dashed red line.
The temperatures of $-4\:{}^{\circ}$C and $+4\:{}^{\circ}$C are indicated by small vertical dashed black lines.
}
\vspace{0mm}
\label{fig_Li}
\end{figure}

The common motivation of all papers by Xingrong Wang since 1995 is to claim that ``the latent heat should be continuous'', and that there is a need to ``eliminate the irrational assumption'' that ``condensation only occurs upon saturation'', with a need to ``remove the discontinuity'' at $0\:{}^{\circ}$C.
It is indeed possible to write the latent heat $L(T)$ for all absolute temperatures by using the discontinuous Heaviside step function $H(T) = 1$ if $T >0\:{}^{\circ}$C and $= 0$ otherwise, leading to
\begin{equation}
L(T) \; = \; L_{fus}(T) \; [ \: 1-H(T) \: ] \: + \:  L_{vap}(T) \: .
\label{eq_L}
\end{equation}
A comparison of Eqs.(\ref{eq_LWH18}) and (\ref{eq_L}) shows that $1-H(T)$ is replaced by $\varphi(T)$ in WH18.
The smooth transition of $L^{\ast}(T)$ between $L_{fus}(T)$ and $L_{vap}(T)$ is indeed visible in Fig.\ref{fig_Li}, where $L^{\ast}(T)$ is plotted from Eq.(\ref{eq_LWH18}).

However, the arguments of Xingrong Wang are based on a misleading interpretation of the latent heat, which cannot be continuous, since the latent heats are precisely defined in thermodynamics as the differences in enthalpies for two different states of water, with obvious discontinuities at $0\:{}^{\circ}$C between liquid water and ice.
This means that the two latent heats $L_{sub}$ and $L_{vap}$ precisely take these discontinuities into account, and they cannot be somehow interpolated by a smooth function like $\varphi(T)$.

Moreover, the use, in WH18, of the temperature of about $+4\:{}^{\circ}$C, at which the density of water is maximum, seems artificial for defining $\varphi(T)$, 
because a latent heat depends only on the difference in enthalpies and is independent of the density of the species. 
The use of this temperature of about $+4\:{}^{\circ}$C corresponds to the strange impacts shown in Fig.\ref{fig_Li}: $L_{sub}$ decreases for $T > -4\:{}^{\circ}$C, $L_{vap}$ increases for $T < 4\:{}^{\circ}$C and $L^{\ast} \approx 2668$~kJ/kg at $0\:{}^{\circ}$C.
All these results disagree with basic results of the thermodynamics of moist air.

 \section{The New Potential Temperature Introduces Discontinuities}
\label{section_pot_temp}
\vspace{-2mm}

\begin{figure}[h]
\centering
\includegraphics[width=0.98\linewidth]{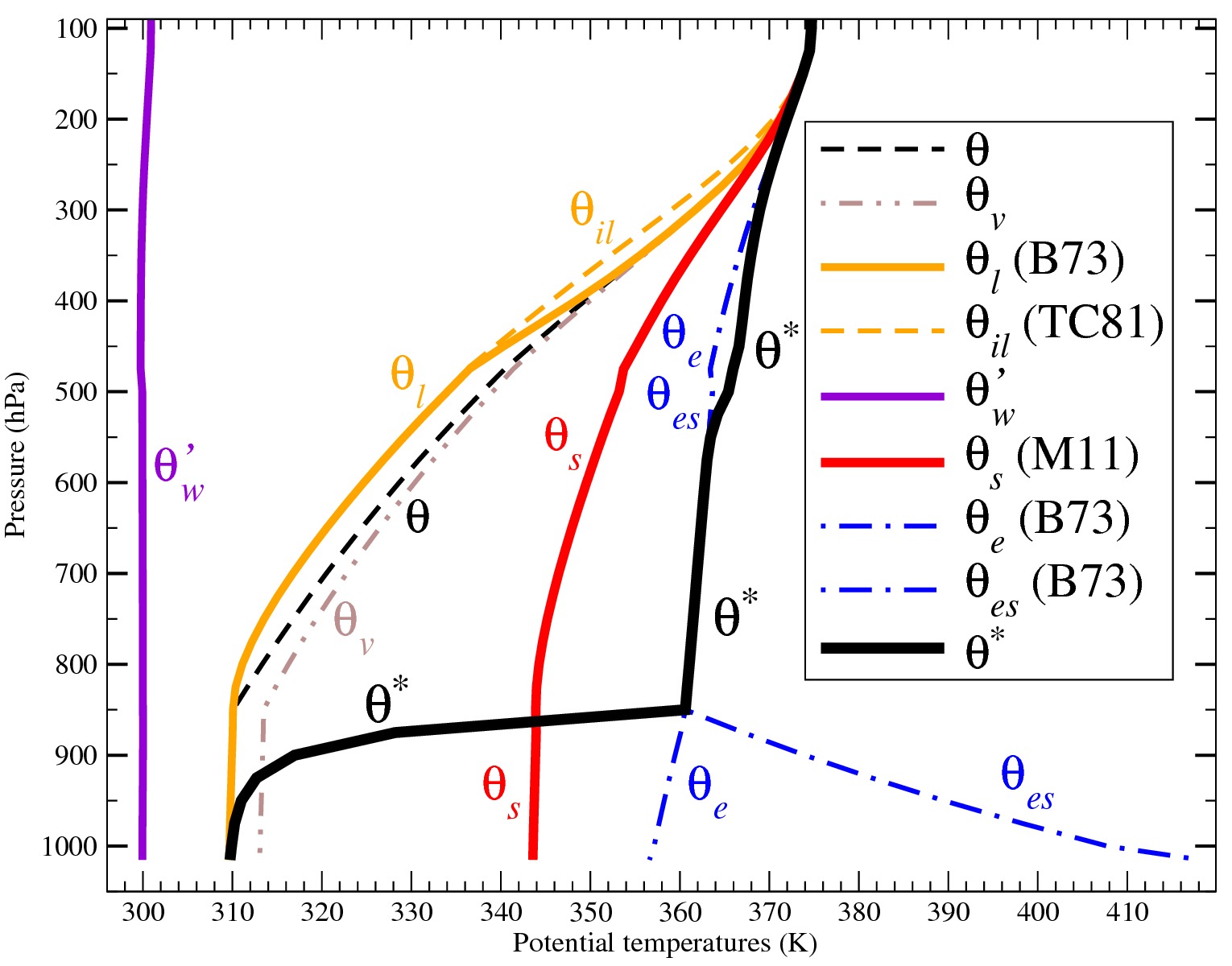}
\vspace{-2mm}
\caption{\small \it
A plot of most known moist-air potential temperatures (in K) for a warm, moist vertical updraught typical of tropical conditions.
The new potential temperature $\theta^{\ast}$ given by Eq.(\ref{equ_theta_star}) is plotted in the bold solid curve.
The surface pressure, absolute temperature, relative humidity, specific water content and saturating value are $p_s = 1015.2$~hPa, $T=311$~K, $RH \: =45$~\%, $q_v = 18.2$~g/kg and $q_{sw} = 41$~g/kg.
The clear-air unsaturated boundary layer extends from the surface up to about  $850$~hPa. 
}
\vspace{0mm}
\label{fig_profiles}
\end{figure}

A new potential temperature is defined in WW18 by:
\begin{equation}
\theta^{\ast} \; = \; \theta \: 
\exp\left[\:
   \frac{L^{\ast}(T) \; q_s(T,p)}
        {c_p \: T} \; 
   {\left(\frac{q_v}{q_s}\right)}^k
\:\right]
\: ,
\label{equ_theta_star}
\end{equation}
where $k=9$, $q_s$ is the saturating value of the specific content $q_v$, and $L^{\ast}(T)$ is given by Eq.~(\ref{eq_LWH18}).
A certain 
updraft of a mixed-phase cumulus is used to compute and plot most of the existing potential temperatures in Fig.\ref{fig_profiles}, including the new one $\theta^{\ast}$.

The saturated equivalent potential temperature $\theta_{es}$ derived in \citet{Betts_73} is a special case of $\theta^{\ast}$ if $L^{\ast} = L_{vap}$ and $k=0$, and the equivalent potential temperature $\theta_e$ can be obtained by replacing the saturating value $q_s$ by the water vapor value $q_v$.
The liquid-water potential temperature $\theta_l$ is defined in \citet{Betts_73}, the virtual potential temperature $\theta_v$ in \citet{Lilly_1968}, the ice-liquid formulation $\theta_{il}$ in \citet{Tripoli_Cotton_81} and the entropy potential temperature $\theta_s$ in \citet{Marquet11,Marquet17}.





All the potential temperatures $\theta'_w$, $\theta$, $\theta_l$, $\theta_{il}$, $\theta_v$ and $\theta_s$ are almost constant in the boundary layer below the condensation level, whereas $\theta_e$ increases slightly with height due to the term $L_{vap}(T)/T$, which increases for decreasing values of $T$ (see Fig.\ref{fig_Li}).
This is not the case for $\theta_{es}$ and $\theta^{\ast}$, which both exhibit intriguing features in the boundary layer.

On the one hand, the saturating value $\theta_{es}$ decreases very rapidly with height, by about $-50$~K between $p_s$ and $850$~hPa, due to the large decrease of $q_s(T,p)$ with $T$.
On the other hand, the new potential temperature $\theta^{\ast}$ increases rapidly with height, by about $+50$~K, for this very moist and warm tropical profile.
It follows that $\theta^{\ast}$ gradually switches from the small, surface values of 
$\theta = \theta_l = \theta_{il} \approx \theta_v$
toward the large values of 
$\theta_e = \theta_{es}$
at the condensation level.
This strange feature observed for $\theta^{\ast}$ is a consequence of $(q/q_s)^9 \approx 1$ at the condensation level, thus with $\theta^{\ast} \approx \theta_{es}$, and with $(q/q_s)^9 \approx 0$ for lower relative humidities, thus with $\theta^{\ast} \approx \theta \exp(0) = \theta$ close to the surface.
 
The potential temperatures $\theta'_w$, $\theta_v$ and $\theta_s$ possess clear physical meanings, since they correspond to pseudo-adiabatic processes, buoyancy force and entropy state function, respectively.
Similarly, $\theta_l$ and $\theta_e$ are adiabatic quantities that are conserved if $q_t$ is a constant.
In contrast, the switch of $\theta^{\ast}$ between $\theta \approx \theta_l$ and $\theta_{es}$ is proof that $\theta^{\ast}$ cannot be related to any of these properties, and cannot have a clear physical meaning.

Moreover, this rapid increase of  
$(q/q_s)^9$ and $\theta^{\ast}$ 
in the boundary layer corresponds to a large vertical gradient 
$\partial \theta^{\ast}/\partial z$ 
and, therefore, must have large impact on the values of the potential vorticity
defined by 
$P_m = (\rho)^{-1} \; \vec{\zeta}_a \: . \: \vec{\nabla} \theta^{\ast}
\approx
(\rho)^{-1}  \: (\zeta + f) \: \partial \theta^{\ast}/\partial z$ in WH18.
This large impact of the term  $(q/q_s)^9$ on the definition of $P_m$ and $P_0 = \rho \: P_m$ in the boundary layer is neither shown nor discussed in 
WH18.

Lastly, the impact of $K = (q/q_s)^9$ at the saturation level leads to a paradoxical issue:
$K$ is introduced in WH18 to ``remove certain discontinuities'', although 
there is clearly a new discontinuous change in the gradient 
$\partial \theta^{\ast}/\partial z$ at the condensation level in Fig.\ref{fig_profiles}.
This discontinuity of the derivative of $K$ at $q=q_s$ is obvious in Fig.\ref{fig_HU} at $RH = 100$~\% and is explained by the derivative $K'(1-\varepsilon) \approx 9$ and $K'(1+\varepsilon) =0$ for small $\varepsilon$.
This strange feature is shared only with $\theta_{es}$ and must greatly impact both $P_m$ and $P_0$.
This discontinuity can be removed by choosing the arbitrary function $G(q/q_s)$  invented for this Comment, plotted in Fig.\ref{fig_HU} and given by Eq.(\ref{eq_GX}) with $G'(1-\varepsilon)\approx 0$.


\bibliographystyle{ametsoc2014}
\bibliography{Marquet_JGR_Atm_arXiv}

           \newpage  

\setcounter{equation}{0} 

\begin{center} 
{\Large 
{\bf Appendix~1:~}{\it A Translation of the Chinese paper\/}\\
\vspace{2mm}
{\bf Wang} Xingrong${}^{1}$ and {\bf Wu} Kejun${}^{1}$. (1995)\\
{\bf 
\vspace{1mm}
{\color{blue}
Discussions on several problems in wet air dynamics.
}\\
\vspace{1mm}
{\color{black}
Or the Authors (official?) title:
}\\
{\color{blue}
An approach to the dynamics of moist air.
}\\
\vspace{2mm}
Scientia Meteorologica Sinica, 1995, 15(1), 9-17.
}
}
\end{center}
\vspace{0 mm}

(1) Anhui Institute of Meteorological Sciences, Hefei, 230061.
\vspace{3 mm}

Received on August 1st, 1994, received revised draft on November 25th
\vspace{3 mm}

{\noindent \small WWW link:
\url{https://caod.oriprobe.com/issues/53446/toc.htm}
}

{\noindent \small WWW link:
\url{https://caod.oriprobe.com/order.htm?id=1477680&ftext=base}
}

{\noindent \small The PDF: $\;$
\url{http://www.jms1980.com/ch/reader/create_pdf.aspx?file_no=19950102&year_id=1995&quarter_id=1&falg=1}
}

\vspace*{4mm}
{\bf \large \underline{Abstract}}
\vspace*{2mm}
\label{Summary}

Based on a certain observational fact, this paper establishes a $P$-coordinate dynamics equation group based on the non-static equilibrium parameter that does not necessarily saturate the wet air and the static balance by introducing the non-hydrostatic balance parameter of the condensing probability function.

According to this equation group, the following conclusions are obtained:
\begin{enumerate}
\item
The wet balance is leveled on the equipotential line, and the wind speed is proportional to the potential gradient and the non-static balance parameter, which is generally smaller than the ground rotation;
\item
The super-earth transfer of the rapids in the rainstorm area is not a quasi-balance of saturated and humid air, but an imbalance related to the rapid increase and weakening of the rapids;
\item
Under the condition of balance, the variation of wind speed with height can be calculated by the wet heat wind equation. When the intersection angle between the ground wind and the hot wind direction does not exceed 90 degrees, the wind speed of the wet heat wind becomes larger than the measured wind speed.
\end{enumerate}

\vspace*{4mm}
{\bf \large \underline{I - Introduction}}
\vspace*{2mm}
\label{I_Intro}

Water vapor is closely related to the activity of the sky.

With the deepening of the analysis and forecasting of typhoons and heavy rains, people pay more and more attention to the effect of water vapor on the occurrence and development of weather systems, and carry out research on wet air dynamics, trying to make a theory of the role of water vapor from the perspective of dynamics analysis.
The existing method is based on the law of conservation of energy in wet air (Xie Yibing, 1980). 

It is recommended to replace the isotherm with the equal total wet temperature 
\begin{equation}
\nonumber
\hspace{20mm}
T_e \; = \;
T \: + \: \frac{g \: Z}{C_p} \: + \: \frac{L \: q_s}{C_p}
\; \approx \; \mbox{Constant}
\: ,
\hspace{20mm} (1)
\end{equation}
or the equivalent potential temperature
\begin{equation}
\nonumber
\hspace{20mm}
\theta_{se} \; = \;
\theta \: \exp\left(
 \frac{L \: q_s}{C_p \: T}
           \right)
\; \approx \; \mbox{Constant}
\:
\hspace{20mm} (2)
\end{equation}
on the isobaric maps.

And based on this, a series of characteristics of wet air movement are discussed [1-2].

Since then, this treatment method has been further extended to storm research (Wang Liangming, 1980), and emphasizes that the atmospheric condensation process can be regarded as a generalized thermal process inside a saturated humid air system. 
Condensation latent heat can be regarded as part of the potential energy of the total system, if the wet air is regarded as a dry atmosphere. 
The generalized temperature 
\begin{equation}
\nonumber
\hspace{40mm}
T^{\ast} \; = \;
T \: + \: \frac{L \: q}{C_p}
\:
\hspace{40mm} (3)
\end{equation}
and the generalized potential
\begin{equation}
\nonumber
\hspace{20mm}
\varphi^{\ast} 
\; = \;
\int_0^{Z*} \: g \: dz
\; = \;
- \: \int_{p_0}^p \: R_d \: T^{\ast} d\ln(p)
\:
\hspace{20mm} (4)
\end{equation}
are intended to extend the equilibrium process of dry air and hot wind into saturated humid air, with the the so-called wet balance wind and the pseudo-balance of the warm wind.

It is considered that the uneven distribution of the wetness in the saturated wet air reflects the uneven distribution of the wet baking, which causes the baroclinicity of the saturated wet air, which causes some changes in the basic characteristics of the dynamic mechanics [3-5].

In saturated humid air, due to the condensation of water vapor, the atmosphere is heated and the static equilibrium is destroyed to some extent. The current fairly dry atmospheric dynamics replace the non-hydrostatic equilibrium equation, with the following general static equilibrium equation \begin{equation}
\nonumber
\hspace{40mm}
\frac{\partial \, \Phi}{\partial P}
\; = \;
- \: \frac{1}{\rho^\ast}
\: \hspace{5mm} \mbox{(see: [2])}
\hspace{40mm} (5)
\end{equation}
or 
\begin{equation}
\nonumber
\hspace{40mm}
\frac{\partial P}{\partial Z}
\; = \;
- \: \rho^\ast \: g
\: \hspace{5mm} \mbox{(see: [4])}
\hspace{40mm} (6)
\end{equation}
respectively, and with corresponding changes in other equations.

However, there are two obvious shortcomings in the wet air dynamics model described above:
\begin{enumerate}
\item It requires that the whole atmosphere is always saturated. In the actual large-scale movement, there is a downdraft in the updraft. 
In the sinking area, the initial saturated air also becomes unsaturated [6-7]. 
Therefore, it is reasonable to describe the actual wet air problem.
\item In the wet process, it is worthwhile to use the wide static balance to replace the non-static balance.
\end{enumerate}

In view of these two points, based on certain observational facts, this paper proposes a scheme to remove the strict condition that the whole atmosphere is always saturated in the wet air kinetics, and establishes a set of processes that are not necessarily saturated and contain condensation and evaporation. Wet aerodynamic basic equations.
On this basis, the $P$-coordinate transformation system obtained by the coordinate conversion is not extracted by static balance, and the wet air aerodynamics equation group in the P coordinate is derived.
According to this equation group, the properties of wet air and hot wind are discussed, and compared with the results of relatively dry atmospheric dynamics.

\vspace*{4mm}
{\bf \large \underline{II - No saturation, wet air, aerodynamics, basic equations}}
\vspace*{2mm}
\label{II_Equations}

It is not difficult to find out the cause of the first defect of the current wet air aerodynamics model. 
This defect is mainly caused by the understanding and treatment of the water vapor condensation process by the current dynamic meteorology.
Therefore, in the current popular numerical forecasting work, the important role of water vapor is usually considered by introducing condensation heating and water vapor equations. 
One of the important assumptions is that the condensation process starts at a certain critical relative value. 
This assumption forces the introduction of a discontinuous $\delta$ function in the water vapor equation.
$Z$ is due to the discontinuity of the $\delta$ function, making the dynamic analysis of the wet aerodynamics at the boundary between the condensation zone and the non-condensation zone extremely difficult, forcing the wet aerodynamic mode to require that the gas is always maintained saturated throughout the day. Reasonable requirements.
In order to remove the whole atmosphere in the wet air aerodynamics, it is always a strict condition to maintain saturation. Let us revisit the related cloud and physical research [8].

According to the experiment, Aitken pointed out: ``When water vapor condenses in the atmosphere, it always adheres to the hygroscopic and non-hygroscopic particles present in the atmosphere, or it acts as a condensation step due to the capture of small droplets of solution, or trace impurities. The surface is made of a hygroscopic surface and most of it becomes hygroscopic particles.'' 

Therefore:
\begin{enumerate}
\item For the atmosphere containing a certain amount of water vapor, there is always a part of the water vapor involved in the condensation evaporation process, so that a certain amount of liquid water can be maintained in the atmosphere. The water vapor and liquid water conservation equations can be written separately:
\begin{equation}
\nonumber
\hspace{10mm}
 \frac{dq}{dt} 
\; = \;
  \frac{d}{dt} \left( K \: q_s \right)
\: ,
\hspace{20mm} (7) 
\end{equation}
\begin{equation}
\nonumber
\hspace{10mm}
   - \:
  \frac{d}{dt} \left( K \: q_s \right)
\; = \;
\frac{dm}{dt} 
\: + \: \eta 
\: ,
\hspace{20mm} (8) 
\end{equation}
where $K$ is the probability function of the water vapor involved in the condensation evaporation process, $m$ is the total amount of liquid water, and $\eta$ is the sinking rate of the liquid water.

\item Under the same atmospheric conditions (gas pressure $P$, temperature $T$, specific humidity $q$ and hygroscopic particle state function $N$), the total liquid water $m$ and the condensing probability function $K$ in the unit volume are the same, that is:
\begin{equation}
\nonumber
\hspace{10mm}
m \; = \;
m \left(
p, T, q, N
\right)
\hspace{20mm} (9) 
\end{equation}
\begin{equation}
\nonumber
\hspace{10mm}
K \; = \;
K \left(
p, T, q, N
\right)
\hspace{20mm} (10) 
\end{equation}
\end{enumerate}

\noindent
Obviously, for the whole atmosphere and since all the air in the unit volume is filled with wet air, there is:
\begin{equation}
\nonumber
\hspace{10mm}
K \; = \;
K \left(
p, T, q, N
\right) \; = \; 1
\hspace{10mm} \mbox{when}\;  q = q_s
\: ,
\hspace{20mm} (11) 
\end{equation}
and if the air that is completely dry, there is
\begin{equation}
\nonumber
\hspace{10mm}
K \; = \;
K \left(
p, T, 0, N
\right) \; = \; 0
\hspace{10mm} \mbox{when}\;  q = 0
\: .
\hspace{20mm} (12) 
\end{equation}
According to the characteristic of $K$ reflected by (11-12), the function of $K$ can be written as
\begin{equation}
\nonumber
\hspace{0mm}
K
\; = \;
{\left(
q/q_s
\right)}^{\alpha}
\: ,
\hspace{20mm} (13) 
\end{equation}
where $\alpha$ is the $Z$ real function of $P, T, q, N$.

If the function of $m$ and $\alpha$ is assumed to be known, and if the relationship between $N$ and $\eta$ can be given by
\begin{equation}
\nonumber
\hspace{10mm}
 \frac{dN}{dt} 
\; = \;
  f_1 \left( 
    p, T, q, m, N, \eta
  \right)
\: ,
\hspace{20mm} (14) 
\end{equation}
then according to the above physical vision, and using the existing dynamics work, we can derive the more rigorous frictionless, wet adiabatic, non-saturated basic set of equations (15) for wet aerodynamics:
\begin{equation}
\nonumber
\hspace{0mm}
\frac{d \, \vec{v}}{dt}
\: + f \: \vec{k} \times  \vec{v} 
\; = \;
 - \: \frac{1}{\rho} \: \nabla P
\: ,
\hspace{20mm} (15) 
\end{equation}
\begin{equation}
\nonumber
\hspace{0mm}
\frac{d \, w}{dt}
\; = \;
 - \: \frac{1}{\rho} 
   \: \frac{\partial \, p}{\partial z}
\: - \: g
\: ,
\hspace{20mm} (15) 
\end{equation}
\begin{equation}
\nonumber
\hspace{0mm}
\frac{d \, T}{dt}
\: - \: \frac{1}{C_p} \: \frac{1}{\rho} 
   \: \frac{d \, P}{d t}
\; = \;
 - \: \frac{L}{C_p} \;
 \frac{
 d \left[ \:
 {(q/q_s)}^{\alpha} \: q_s
 \: \right]
 }{dt} 
\: ,
\hspace{20mm} (15) 
\end{equation}
\begin{equation}
\nonumber
\hspace{0mm}
\frac{1}{\rho} \:
\frac{d \, \rho}{dt}
\: + \:
\nabla \, . \, \vec{v}
\: + \:
\frac{\partial \, \omega}{\partial z}
\; = \;
 0
\: ,
\hspace{20mm} (15) 
\end{equation}
\begin{equation}
\nonumber
\hspace{0mm}
P
\; = \;
 \rho \: R \: T
\: ,
\hspace{20mm} (15) 
\end{equation}
\begin{equation}
\nonumber
\hspace{0mm}
\frac{d \, q}{dt}
\; = \;
 \frac{
 d \left[ \:
 {(q/q_s)}^{\alpha} \: q_s
 \: \right]
 }{dt} 
\: ,
\hspace{20mm} (15) 
\end{equation}
\begin{equation}
\nonumber
\hspace{0mm}
- \:
 \frac{
 d \left[ \:
 {(q/q_s)}^{\alpha} \: q_s
 \: \right]
 }{dt} 
\; = \;
\frac{d \, m}{dt}
\: + \: \eta
\: ,
\hspace{20mm} (15) 
\end{equation}
\begin{equation}
\nonumber
\hspace{0mm}
\frac{d \, N}{dt}
\; = \;
  f_1 \left( 
    p, T, q, m, N, \eta
  \right)
\: .
\hspace{20mm} (15) 
\end{equation}
For such a system of equations, although the physical model is more rigorous, due to the lack of basic understanding of $\alpha$, $m$ and $f_1$, it is very difficult to deal with practical problems in strict accordance with Equation~14.

However, if $\alpha$ is assumed to be a function of $P, T, q$, and for a specific problem, $\alpha$ can be determined by experiments and a large amount of data analysis, such as $q_s$, then you can bypass $m$, $N$ in the discussion with $\eta$, and the equations (15) are simplified into (16):
\begin{equation}
\nonumber
\hspace{0mm}
\frac{d \, \vec{v}}{dt}
\: + f \: \vec{k} \times  \vec{v} 
\; = \;
 - \: \frac{1}{\rho} \: \nabla P
\: ,
\hspace{20mm} (16) 
\end{equation}
\begin{equation}
\nonumber
\hspace{0mm}
\frac{d \, w}{dt}
\; = \;
 - \: \frac{1}{\rho} 
   \: \frac{\partial \, p}{\partial z}
\: - \: g
\: ,
\hspace{20mm} (16) 
\end{equation}
\begin{equation}
\nonumber
\hspace{0mm}
\frac{d \, T}{dt}
\: - \: \frac{1}{C_p} \: \frac{1}{\rho} 
   \: \frac{d \, P}{d t}
\; = \;
 \frac{L}{C_p} \;
 \frac{
 d \left[ \:
 {(q/q_s)}^{\alpha(p,T,q)} \: q_s(P,T)
 \: \right]
 }{dt} 
\: ,
\hspace{20mm} (16) 
\end{equation}
\begin{equation}
\nonumber
\hspace{0mm}
P
\; = \;
 \rho \: R \: T
\: ,
\hspace{20mm} (16) 
\end{equation}
\begin{equation}
\nonumber
\hspace{0mm}
\frac{1}{\rho} \:
\frac{d \, \rho}{dt}
\: + \:
\nabla \, . \, \vec{v}
\: + \:
\frac{\partial \, \omega}{\partial z}
\; = \;
 0
\: ,
\hspace{20mm} (16) 
\end{equation}
\begin{equation}
\nonumber
\hspace{0mm}
\frac{d \, q}{dt}
\; = \;
 \frac{
 d \left[ \:
 {(q/q_s)}^{\alpha(P,T,q)} \: q_s(p,T)
 \: \right]
 }{dt} 
\: ,
\hspace{20mm} (16) 
\end{equation}

\vspace*{1mm} \noindent
These are the basic equations of the general aerodynamics of the $Z$-coordinate simplification of the non-viscous wet adiabatic.
It can be seen from this group (16) that it is suitable for atmospheric motion with various humidity states.
In the ideal state, when $q$ is $q=0$ and $q=q_s$ respectively, they are degraded into the dry atmospheric dynamics equation group and the saturated wet air dynamics group of equations given by the literature (see [1]).
And because we use continuous and unsatisfactory condensing and evaporating processes, the probability function of water vapor replaces the unsuccessful delta function, eliminating the previous delta function, which makes it difficult to perform dynamic analysis at the boundary of the condensation zone.
This basically compensates for the first (1) defect in the current wet air dynamics model indicated in the introduction:

\vspace*{4mm}
{\bf \large \underline{III - The P-coordinate wet air equation without static balance}}
\vspace*{2mm}
\label{III_Equations}

It is necessary to discuss the deviations of the dynamics equations that the relatively dry atmospheric dynamics equations [4,7] and the wet air are actually observing, and only need to compare the two.
Since the relatively dry and wet aerodynamic modes that are currently available are based on the general static equilibrium, the equivalent dry equations of the $Z$ coordinate are given by the dry air dynamics equations, which is premised on the general static balance.
The vertical coordinate transformation formula is transformed into the $P$ coordinate equation.
Therefore, if it is possible to strictly derive the $P$ coordinate conversion formula that is not based on the static balance.
By converting the equation (16) to the $P$ coordinate form, it is possible to discuss the deviation of the relatively dry atmospheric dynamics equation group.

Analyzing the derivation of the vertical coordinate conversion in the general book of education [9], it can be seen that the static horizontal balance formula 
\begin{equation}
\nonumber
\hspace{0mm}
\alpha \; \frac{\partial p}{\partial z}
\; = \;
 - \: g
\: ,
\hspace{20mm} (17) 
\end{equation}
has the main horizontal pressure gradient force and continuous equation conversion formula.

Note that the vertical motion equation
\begin{equation}
\nonumber
\hspace{0mm}
\frac{dw}{dt}
\; = \;
 - \: \frac{1}{\rho} \; \frac{\partial p}{\partial z}
 - \: g
\: ,
\hspace{20mm} (18) 
\end{equation}
can be rewritten as
\begin{equation}
\nonumber
\hspace{0mm}
\left(
1 \: + \: 
\frac{1}{g} 
\frac{dw}{dt}
\right)
\; = \;
- \:
\frac{R \: T}{p \: (\partial \Phi / \partial p)}
\; = \;
 - \: \frac{1}{\rho \: g} \; \frac{\partial p}{\partial z}
\: ,
\hspace{20mm} (19) 
\end{equation}
to define 
\begin{equation}
\nonumber
\hspace{0mm}
\varepsilon
\; = \;
1 \: + \: 
\frac{1}{g} 
\frac{dw}{dt}
\: ,
\hspace{20mm} (20) 
\end{equation}
as the non-hydrostatic equilibrium parameter. 
Using (19), the expression for the horizontal pressure gradient force in any $\xi$ coordinate, and the continuous equation in $\xi$ coordinate without making the static balance, can be written as
\begin{equation}
\nonumber
\hspace{0mm}
- \: \alpha \: \nabla_h \, p
\; = \;
   - \: \alpha \: \nabla_\xi \, p
\: + \:
 \alpha \; \frac{\partial p}{\partial z}
        \; \nabla_\xi \, z
\; = \;
   - \: \alpha \: \nabla_\xi \, p
\: - \: \varepsilon \:
        \: \nabla_\xi \, \varphi
\: ,
\hspace{10mm} (21) 
\end{equation}
\begin{equation}
\nonumber
\hspace{0mm}
\frac{d}{dt} \left( \ln \rho \right)
\: + \:
\nabla_h \, . \, \vec{V}
\: + \:
\frac{\partial \, \omega}{\partial z}
\:
\hspace{89mm} (22) 
\end{equation}
\begin{equation}
\nonumber
\hspace{32mm}
\; = \;
\frac{d}{dt} \left( \ln \rho \right)
\: + \:
\frac{d}{dt} 
\left[\: 
   L \: \left(
     \frac{\partial \, z}{\partial \xi}
       \right)
\: \right]
\: + \:
\nabla_\xi \, . \, \vec{V}
\: + \:
\frac{\partial \, \xi}{\partial \xi}
\: ,
\hspace{15mm} (22) 
\end{equation}
\begin{equation}
\nonumber
\hspace{32mm}
\; = \;
\frac{d}{dt} \left[
   \ln 
 \left( \rho \:
     \frac{\partial \, z}{\partial \xi}
 \right)
\right]
\: + \:
\nabla_\xi \, . \, \vec{V}
\: + \:
\frac{\partial \, \xi}{\partial \xi}
\: ,
\hspace{35mm} (22) 
\end{equation}
\begin{equation}
\nonumber
\hspace{32mm}
\; = \;
\frac{d}{dt} \left[
   \ln 
 \left( \frac{1}{\varepsilon} \:
     \frac{\partial \, P}{\partial \xi}
 \right)
\right]
\: + \:
\nabla_\xi \, . \, \vec{V}
\: + \:
\frac{\partial \, \xi}{\partial \xi}
\: .
\hspace{35mm} (22) 
\end{equation}
Let us then set $\xi=P$, then convert the equations (16) to $P$ coordinate equations according to (21-22), and use other vertical coordinate transformation relations in the general textbook that are not based on static balance, to arrive at
\begin{equation}
\nonumber
\hspace{0mm}
\frac{d \, \vec{V}}{dt}
\: + f \: \vec{k} \times  \vec{V} 
\; = \;
 \varepsilon \: \nabla_P \, \Phi
\: \; ,
\hspace{20mm} (23) 
\end{equation}
\begin{equation}
\nonumber
\hspace{0mm}
\frac{d \, T}{dt}
\: - \: \frac{R}{C_p} \: \frac{T}{P} 
   \: \omega
\; = \;
 \frac{L}{C_p} \;
 \frac{
 d \left[ \:
 {(q/q_s)}^{\alpha(p,T,q)} \: q_s(P,T)
 \: \right]
 }{dt} 
\: \; ,
\hspace{20mm} (23) 
\end{equation}
\begin{equation}
\nonumber
\hspace{0mm}
\frac{d}{dt}
\left[ \:
 \ln\left( \frac{1}{\varepsilon} \right)
\: \right]
\: + \: 
\frac{\partial u}{\partial x}
\: + \: 
\frac{\partial v}{\partial y}
\: + \: 
\frac{\partial \omega}{\partial p}
\; = \; 0
\: \; ,
\hspace{20mm} (23) 
\end{equation}
\begin{equation}
\nonumber
\hspace{0mm}
\frac{d^2}{dt^2} \: \Phi
\; = \;
 \left( \varepsilon - 1 \right) \: g^2
\: ,
\hspace{20mm} (23) 
\end{equation}
\begin{equation}
\nonumber
\hspace{0mm}
\varepsilon
\; = \;
 - \: \frac{R \: T}{P \: (\partial \Phi / \partial P)}
\: ,
\hspace{20mm} (23) 
\end{equation}
\begin{equation}
\nonumber
\hspace{0mm}
\frac{d \, q}{dt}
\; = \;
 \frac{
 d \left[ \:
 {(q/q_s)}^{\alpha(P,T,q)} \: q_s(p,T)
 \: \right]
 }{dt} 
\: ,
\hspace{20mm} (23) 
\end{equation}

\vspace*{4mm}
{\bf \large \underline{IV - The equivalent dry atmospheric dynamics equations}}
\vspace*{2mm}
\label{IV_Equations}

With the above equations, from references [2] and [4], and
considering the atmospheric condensation process as a generalized thermal process inside the wet air, that is, considering the latent heat of condensation as part of the total energy of the system, the set of Equations~(23) can be transformed into fairly dry atmospheric dynamics equations.

The equation
\begin{equation}
\nonumber
\hspace{0mm}
\frac{d \, T}{dt}
\: - \: \frac{R}{C_p} \: \frac{T}{P} 
   \: \omega
\; = \;
 - \: \frac{L}{C_p} \;
 \frac{
 d}{dt} \left[ \:
 {\left(\frac{q}{q_s}\right)}^{\alpha(p,T,q)} \: q_s(P,T)
 \: \right]
\: \; ,
\hspace{20mm} (24) 
\end{equation}
obtained from the thermodynamic equation
\begin{equation}
\nonumber
\hspace{0mm}
 \frac{d}{dt} \ln
 \left(
  T \: P ^{-R/C_p}
 \right)
\; = \;
 - \: \frac{L}{C_p} \;
   \: \frac{1}{T} \;
 \frac{d}{dt} \left[ \:
 {\left(\frac{q}{q_s}\right)}^{\alpha(p,T,q)} \: q_s(P,T)
 \: \right]
\: \; ,
\hspace{20mm} (25) 
\end{equation}
can be proved in general from [9]
\begin{equation}
\nonumber
\hspace{0mm}
  O \left\{ \;
 \frac{d}{dt} \left[ \:
 {\left(\frac{q}{q_s}\right)}^{\alpha(p,T,q)} \: q_s(P,T)
 \: \right]
 \; \Big/ \;
 \left[ \:
 {\left(\frac{q}{q_s}\right)}^{\alpha(p,T,q)} \: q_s(P,T)
 \: \right]
 \: \right\}
\; \gg \;
  O \left\{ \:
   \frac{1}{T} \frac{dT}{dt}
 \: \right\}
\: \; ,
\hspace{5mm} (26) 
\end{equation}
Therefore, (25) can be approximated by
\begin{equation}
\nonumber
\hspace{0mm}
 \frac{d}{dt} \ln
 \left(
  T \: P ^{-R/C_p}
 \right)
\: + \:
 \frac{d}{dt} \left[ \:
      \frac{L}{C_p} \;
   \: \frac{q_s}{T} \;
 {\left(\frac{q}{q_s}\right)}^{\alpha(p,T,q)}
 \: \right]
\; = \;
 0
\: \; ,
\hspace{20mm} (27) 
\end{equation}
i.e.
\begin{equation}
\nonumber
\hspace{0mm}
 \frac{d}{dt} 
 \ln
 \left\{ \;
  T \: P ^{-R/C_p} \:
  \exp
  \left[ \:
      \frac{L}{C_p} \;
   \: \frac{q_s}{T} \;
     {\left(
       \frac{q}{q_s}
     \right)}^{\alpha(p,T,q)}  
  \: \right]
 \; \right\}
\; = \;
 0
\: \; .
\hspace{20mm} (28) 
\end{equation}
In order to correspond to the dry aerodynamic equations, we thus define the generalized temperature
\begin{equation}
\nonumber
\hspace{0mm}
 T^{\ast}
 \; = \;
  T \; \:
  \exp
  \left[ \:
      \frac{L}{C_p} \;
   \: \frac{q_s}{T} \;
     {\left(
       \frac{q}{q_s}
     \right)}^{\alpha(p,T,q)}  
  \: \right]
\: \; .
\hspace{20mm} (29) 
\end{equation}
The equilibrium is established if a generalized static force is assumed. 
It can be seen, from the non-hydrostatic equilibrium parameter equation, that
\begin{equation}
\nonumber
\hspace{0mm}
\frac{R \: T^{\ast}}
     {p \: (\partial \Phi / \partial p)}
\; = \;
 - \: 1
\:
\hspace{20mm} (30) 
\end{equation}
is obtained by deriving
\begin{equation}
\nonumber
\hspace{0mm}
\frac{R \: T}
     {p \: (\partial \Phi / \partial p)}
\; = \;
 - \: \varepsilon
\:
\hspace{20mm} (31) 
\end{equation}
from (29), (30) and the Z coordinate vertical motion equation.

The above general static equilibrium assumption is actually assuming non-hydrostatic balance. 
The non-hydrostatic equilibrium parameter is thus completely determined by
\begin{equation}
\nonumber
\hspace{0mm}
\varepsilon
\; = \;
  \exp
  \left[ \: - \:
      \frac{L}{C_p} \;
   \: \frac{q_s}{T} \;
     {\left(
       \frac{q}{q_s}
     \right)}^{\alpha(p,T,q)}  
  \: \right]
\: .
\hspace{20mm} (32) 
\end{equation}
Using this assumption, the equation (23) can be transformed into a fairly dry wet aerodynamic equation
\begin{equation}
\nonumber
\hspace{0mm}
\frac{d \, \vec{V}}{dt}
\: + f \: \vec{k} \times  \vec{V} 
\; = \;
  - \: \exp
  \left[ \: - \:
      \frac{L}{C_p} \;
   \: \frac{q_s}{T} \;
     {\left(
       \frac{q}{q_s}
     \right)}^{\alpha(p,T,q)}  
  \: \right]
 \: \nabla \, \Phi
\: \; ,
\hspace{20mm} (33) 
\end{equation}
\begin{equation}
\nonumber
\hspace{0mm}
\frac{d \, T^{\ast}}{dt}
\: - \: \frac{R}{C_p} \: \frac{T^{\ast}}{P} 
   \: \omega
\; = \;
 0
\: \; ,
\hspace{20mm} (33) 
\end{equation}
\begin{equation}
\nonumber
\hspace{0mm}
\frac{d}{dt}
\left[ \:
      \frac{L}{C_p} \;
   \: \frac{q_s}{T} \;
     {\left(
       \frac{q}{q_s}
     \right)}^{\alpha(p,T,q)} 
\: \right]
\: + \: 
\frac{\partial u}{\partial x}
\: + \: 
\frac{\partial v}{\partial y}
\: + \: 
\frac{\partial \omega}{\partial p}
\; = \; 0
\: \; ,
\hspace{20mm} (33) 
\end{equation}
\begin{equation}
\nonumber
\hspace{0mm}
\frac{d^2}{dt^2} \: \Phi
\; = \;
 \left\{ \; 
  \exp
  \left[ \: - \:
      \frac{L}{C_p} \;
   \: \frac{q_s}{T} \;
     {\left(
       \frac{q}{q_s}
     \right)}^{\alpha(p,T,q)}  
  \: \right]
  \: - \: 1 
  \; \right\} \: g^2
\: ,
\hspace{20mm} (33) 
\end{equation}
\begin{equation}
\nonumber
\hspace{0mm}
\frac{R \: T^{\ast}}{P \: (\partial \Phi / \partial P)}
\; = \;
 - \: 1
\: ,
\hspace{20mm} (33) 
\end{equation}
\begin{equation}
\nonumber
\hspace{0mm}
\frac{d q}{dt}
\; = \;
 \frac{d}{dt}
 \left\{ \: q_s \:
     {\left(
       \frac{q}{q_s}
     \right)}^{\alpha(p,T,q)}  
 \: \right\}
\: ,
\hspace{20mm} (33) 
\end{equation}
which are based on generalized static equilibrium equation (30). 

Combination of (33) with the literature [4] and [7] generates the dry saturated wet air dynamics equations given by the dry-air aerodynamics equations [9]: \\
\begin{equation}
\nonumber
\hspace{0mm}
\frac{d u}{dt}
\; = \;
  f \: v 
\: - \: 
 \frac{\partial \Phi}{\partial x}
\: \; ,
\hspace{20mm} (34) 
\end{equation}
\begin{equation}
\nonumber
\hspace{0mm}
\frac{d v}{dt}
\; = \;
  - \: f \: u 
\: - \: 
 \frac{\partial \Phi}{\partial y}
\: \; ,
\hspace{20mm} (34) 
\end{equation}
\begin{equation}
\nonumber
\hspace{0mm}
 \frac{\partial \Phi}{\partial p}
\; = \;
\: - \: \frac{R \: T^{\ast}}{P}
\: \; ,
\hspace{20mm} (34) 
\end{equation}
\begin{equation}
\nonumber
\hspace{0mm}
\frac{\partial u}{\partial x}
\: + \: 
\frac{\partial v}{\partial y}
\: + \: 
\frac{\partial \omega}{\partial p}
\; = \; 0
\: \; ,
\hspace{20mm} (34) 
\end{equation}
\begin{equation}
\nonumber
\hspace{0mm}
C_p \: 
\frac{d T^{\ast}}{dt}
\: + \:
P \:
 \frac{d}{dt}
 \left( \: 
     \frac{R \: T^{\ast}}{P}
 \: \right)
\; = \; 0
\: .
\hspace{20mm} (34) 
\end{equation}
It can be seen that even in the case of saturation ($q=q_s$), except for the thermodynamic equation and the general static equilibrium equation, the other equations are obviously different.

Because of these differences, the basic characteristics of the actual wet air dynamics equation group and the basic characteristics of the dynamics equations given in the literature [2] and [4] must be significantly different.
The following is a discussion of the issue of local wind and hot wind.

\vspace*{4mm}
{\bf \large \underline{V - Wet air and Earth's wind balanced}}
\vspace*{2mm}
\label{V_Balanced}

{\it ... equations and text in this section to be continued ...}

Under the condition of no horizontal acceleration and no friction, the equilibrium air
\begin{equation}
\nonumber
\hspace{0mm}
\vec{V}_{gh} 
\; = \; 
\dots
\: ,
\hspace{20mm} (35) 
\end{equation}
of the wet air can be obtained from the horizontal motion equation of the equation (33). 

From (35), it can be seen that the dry air is the same as the dry air. 
The direction of the wind is parallel to the equipotential line. 
Unlike the dry air, the magnitude is not only proportional to the potential gradient, but also proportional to the non-hydrostatic balance parameter 
$\exp(\ldots)$, which is equal to the ground rotation of dry air wind multiplied by non-hydrostatic balance parameters.

Obviously, when $q=0$, the equation (35) is degenerated into a dry air-to-wind relationship.

??? ... (35) ... ??? ... $\exp(\ldots)$ is a positive real number smaller than this, so that the wet balance wind should be smaller than the ground.
This seems to indicate that the release of latent heat from water vapor condensation is not the direct cause of the low - altitude jet super - geostrophic feature.
Speculation in the literature (4): ``The low-altitude jets in the rainstorm areas that are usually observed are in the same direction as the ground, but the so-called super-geostrophic phenomenon with a large wind speed may be considered after the generalized potential $\varphi *$ The quasi-equilibrium phenomenon of saturated humid air''. 
It is only a false guess...

In fact, the analysis of the characteristics of the low-altitude super-geo-rotation (10) points out that for low-level jets, ``not all moments are super-geost, and when the jets are enhanced, both $u$ and $u_g$ are increasing, $u>u_g$ (That is, super-transfer); when 12 hours after $u$ reaches the maximum value, $u_g$ reaches the maximum value, $u<ug$, but in the process of $u_g$ reduction, it turns into $u>u_g$.''

This clearly shows that the super-geostation phenomenon is an imbalance that cannot be ignored by the horizontal force velocity $du/dt$ associated with the strength and attenuation of the jet force; a period of time after the rapid flow that can be regarded as no horizontal acceleration reaches the maximum value The measured wind is less than the floor tile wind.
A conclusion is consistent with (35).

Will use the formula 35 for $P$ to get the quotient
\begin{equation}
\nonumber
\hspace{0mm}
\vec{V}_{Th} 
\; = \; 
\dots
\: ,
\hspace{20mm} (36) 
\end{equation}
If defined
\begin{equation}
\nonumber
\hspace{0mm}
T_{\ast} \; {\left( \frac{P}{P_0} \right)}^{-R/Cp} 
\; = \; 
\theta_{se}
\: ,
\hspace{20mm} (37) 
\end{equation}
Then, using the state equations $P = \rho \: k \: T$ and (29), the equation (36) can be rewritten as
\begin{equation}
\nonumber
\hspace{0mm}
\vec{V}_{Th} 
\; = \; 
\dots
\: ,
\hspace{20mm} (38) 
\end{equation}

... in case $exp(\ldots)$ is a negative value and the absolute value is much less than 1, ... the end from (38) is not difficult to find, as long as the wind turns, ... with ... the same magnitude, under balanced wind conditions.
The variation of wind speed with height is approximated by the equation of wet heat and wind.
\begin{equation}
\nonumber
\hspace{0mm}
\vec{V}_{Th} 
\; = \; 
\dots
\: ,
\hspace{20mm} (39) 
\end{equation}

To calculate, and only if the wind and the hot wind direction are the same, the angle of intersection is not more than 90 degrees. The calculated value is always larger than the measured value, and the error is:
\begin{equation}
\nonumber
\hspace{0mm}
\vec{V}_{Th} 
\; = \; 
\dots
\: ,
\hspace{20mm} (40) 
\end{equation}
This result is consistent with the variation of the wind speed with height and the measured wind speed with height variation as indicated by the equation (39) in the case of saturation, as pointed out in the Journal of Meteorology, Vol. 41, No. 2, "Reader's Letter" [11].

\vspace*{4mm}
{\bf \large \underline{VI - Conclusion}}
\vspace*{2mm}
\label{VI_Conclusion}

Based on a certain observational reality, this paper proposes a solution to remove the entire atmosphere in the wet air kinetics, which is always full of such a harsh condition, and establishes a wet air kinetic equation group that does not necessarily contain condensation and evaporation processes.
On this basis, according to the most basic relationship of vertical coordinate transformation and the P coordinate transformation relationship obtained from the vertical motion equation without static balance, the derivation of the P coordinate wettable which is not based on the static balance is derived. Aerodynamic equations.
According to this group, it is compared with the existing dry and wet air equations, and the characteristics of wet air and hot wind are discussed.

What needs to be pointed out is that the work of this article is only the beginning of the initial step. There are many questions that need to be further explored.

\vspace*{4mm}

\noindent {\bf Acknowledgement:} 
Yang Dasheng has put forward some valuable opinions on this article, and I would like to thank you.

\vspace*{4mm}

\vspace*{4mm}
{\bf \large \underline{References}}
\vspace*{2mm}
\label{References}

[1] Xie Yibing, 
   The Problem of Wet and Oblique Air and Turbulence, 
   Rainstorm Collection, 1-15, 
   Jilin People's Publishing House, 1978

[2] Xie Yibing, 
   Steady and Unstable Oblique Pressing, 
   Star Image, 39, 1, 44-57, 1981

[3] Wang Erming et al., 
   Exploring some problems in stormy weather and 
   kinetics, 
   Sun Yat-sen University, 1, 1978

[4] Wang Liangming et al., 
   Basic equations and main features of saturated wet 
   air dynamics, 
   Meteorology, 38, 1, 44-50, 1980

[5] Wang Liangming, 
   Basic Characteristics of Saturation and Wet Air 
   Thermodynamics, 
   Gas Image, 38, 2, 106-109, 1980

[6] Zhu Baozhen, 
   Comment on "The Problem of Wet and Oblique 
   Atmospheric Forces in the Earth", 
   Journal of Gas Imaging, 44, 1, 118-124, 1986

[7] Zhu Baozhen, 
   On the issue of "Basic Formula of Saturation 
   and Wet Air Dynamics", 
   Journal of Gas Imaging, 44, 3, 378-381, 1986.

[8] Mason, B.J. 
   Cloud Physics, Institute of Atmospheric Physics, 
   Chinese Academy of Sciences, 1-32, 
   Science Publishing House, 1978.

[9] Ye Yizheng, Li Chongyin, Wang Bikui, 
   Dynamic Aerography, 60-62, 
   Science Publishing House, 1988

[10] Zhu Gangen, 
    Low-altitude rapids and heavy rain, 
    gas-based technical materials, 8, 12-18, 1975.

[11] Reader's letter to the editor, 
    Press, Illustrated, 41, 2, 247, 1983.

          \newpage 

\setcounter{equation}{0} 

\begin{center} 
{\Large 
{\bf Appendix~2:~}{\it A (crude and partial) translation  of the Chinese paper\/}\\
\vspace{2mm}
{\bf Wang} Xingrong${}^{1}$, {\bf Wang} Zhongxing${}^{2}$ 
and {\bf Shi} Chunxi${}^{3}$ (1998).\\
{\bf 
\vspace{1mm}
{\color{blue}
On the problem of non-conservation of wet vortex (potential 
vorticity) in atmospheric motion.
}\\
\vspace{2mm}
Scientia Meteorologica Sinica, 1998, 18 (2),135-141.
}
}
\end{center}
\vspace{0 mm}

(1) Anhui Institute of Aeronautical Sciences, Hefei, 230061

(2) Department of Earth and Space Science, University of 
    Science and Technology of China, Third World Science 
    Institute, Department of Geoscience and Astronomy, 
    Advanced Research Center, Hefei, 230026

(3) Anhui Institute of Aeronautics and Astronautics, 
    Hefei, 230061
\vspace{3 mm}

Received date : 1997-09-10 ; Revised draft Date : 1997-11-28
\vspace{3 mm}

{\noindent \small WWW link:
\url{http://en.cnki.com.cn/Article_en/CJFDTotal-QXKX199802004.htm}
}

{\noindent \small WWW link:
\url{https://caod.oriprobe.com/order.htm?id=1961740&ftext=base}
}

{\noindent \small The PDF: $\;$
\url{http://www.jms1980.com/ch/reader/create_pdf.aspx?file_no=19980221&year_id=1998&quarter_id=2&falg=1}
}

\vspace*{4mm}
{\bf \large \underline{Abstract}}
\vspace*{2mm}
\label{Summary}

In this paper, the establishment of the wet vortex equation to extract the non-conservation basic criterion is used to analyze the dynamic characteristics of non-conservation motion, especially the non-adiabatic heating and external force field in the wet neutral layer region.

\vspace*{4mm}
{\bf \large \underline{Introduction}}
\vspace*{2mm}
\label{Intro}

At present, in the work of dynamic meteorological theory, the atmospheric motion based on adiabatic and no external force field has been proved to satisfy the vortex conservation.
For non-adiabatic, the vorticity of the vortex in the case of external force field is not limited, except for the non-adiabatic release of the saturated hot air condensing latent heat release [1-5], but other situations are not discussed.

The main reasons are: Most scholars believe that, on average, the effects of other non-adiabatic and external force fields on atmospheric circulation may be more important than condensation potential, but in the evolution of large-scale free-range atmospheric motion, in addition to condensation outside the release of latent heat, other non-adiabatic and external force fields have relatively small effects on vertical motion [6].
Therefore, it is generally not considered in the analysis of the theory.

However, the statistical facts of the ``sudden outbreak'' and ``transitional change'' of many weather phenomena are related to the tidal force of the sun and the moon [7-11] have shown that for mesoscale and large-scale atmospheric motion, not only the condensation latent heat release of water vapor, but, for example, the evolution of the lunar tidal movement, plays an important role.

In 1991, for example, during the period of the flood season in Anhui Province, the beginning and end of the storm period and the singularity of the intersection point are clearly related to each other (see Table 1).

This leads to such a series of questions: \\
1) Under what conditions non-adiabatic heating and external force fields can play a role in the lifting of medium-scale or large-scale atmospheric movements, and cannot be ignored? \\
2) What are the characteristics of the motion during the atmospheric movements that are used for non-adiabatic heating and external force field lifting?

For these questions, this paper is prepared to discuss from the wet vortex equation.

\vspace*{4mm}
{\bf \large 1)  \underline{Wet position vortex equation}}
\vspace*{2mm}
\label{1_Wet_vortex_equ}

\begin{table}
\caption{\it The singularity of the Moon-Perigean period and the beginning and stopping days of the Anhui rain gush in 1991: {\bf see the Chinese PDF}.
\label{Table1}
}
\end{table}

For non-viscous mesoscale, large-scale non-uniform saturation [2] wet free atmospheric motion (so-called non-uniform saturation means partial saturation and other partial unsaturated), if considering condensation of latent heat release, and other non-adiabatic heating And the external force, according to the literature [12], its equation of motion is:
\begin{equation}
\nonumber
\hspace{0mm}
\frac{d \, V}{dt}
\: + f \: K \times  V 
\; = \;
 - \: \frac{1}{\rho} \: \nabla P
 \: + \: E_a
\: ,
\hspace{20mm} (1) 
\end{equation}
\begin{equation}
\nonumber
\hspace{0mm}
\frac{d \, w}{dt}
\; = \;
 - \: \frac{1}{\rho} 
   \: \frac{\partial \, P}{\partial z}
\: - \: g
\: ,
\hspace{20mm} (2) 
\end{equation}
\begin{equation}
\nonumber
\hspace{0mm}
\frac{d \, T}{dt}
\: - \: \frac{1}{C_p} \: \frac{1}{\rho} 
   \: \frac{d \, P}{d t}
\; = \;
 - \: \frac{L}{C_p} \;
 \frac{
 d \left[ \:
 {(q/q_s)}^{\alpha} \: q_s
 \: \right]
 }{dt} 
 \: + \:
\frac{1}{C_p} \: \frac{\delta Q}{\delta t} 
\: ,
\hspace{20mm} (3) 
\end{equation}
\begin{equation}
\nonumber
\hspace{0mm}
P
\; = \;
 \rho \: R \: T
\: ,
\hspace{20mm} (4) 
\end{equation}
\begin{equation}
\nonumber
\hspace{0mm}
\frac{1}{\rho} \:
\frac{d \, \rho}{dt}
\: + \:
\nabla \, . \, V
\: + \:
\frac{\partial \, \omega}{\partial z}
\; = \;
 0
\: ,
\hspace{20mm} (5) 
\end{equation}
\begin{equation}
\nonumber
\hspace{0mm}
\frac{d \, q}{dt}
\; = \;
 \frac{
 d \left[ \:
 {(q/q_s)}^{\alpha} \: q_s
 \: \right]
 }{dt} 
\: ,
\hspace{20mm} (6) 
\end{equation}
where 
$E_a$ is the horizontal component of the external force field, 
${\delta Q}/{\delta t}$ is the non-adiabatic heating other than the latent heat of condensation, 
$(q/q_s)^{\alpha}$ is the function of the condensation probability, 
where ${\alpha}$ is the real number to be determined by the experiment and a large amount of data.
The other symbols are those commonly used.

In [12], the effect of the large-scale impact on the vertical motion during the large-scale and free-range atmospheric motion is mainly caused by the latent heat of condensation.

Therefore, it can be assumed that if there is damage to the static balance, it is mainly caused by the release of latent heat from the condensation of water vapor, that is, if the following temperature is defined:
\begin{equation}
\nonumber
\hspace{0mm}
T^{\ast}
\; = \;
T \;
 \exp
 \left[ \:
 \frac{L}{C_p} \:
 \frac{q_s}{T} \:
 {\left( \frac{q}{q_s} \right)}^{\alpha}
 \: \right]
\: ,
\hspace{20mm} (7) 
\end{equation}
Then there is a broad sense of static balance
\begin{equation}
\nonumber
\hspace{0mm}
\frac{R \; T^{\ast}}
{P \: (\partial \Phi / \partial P)}
\; = \;
-1
\: ,
\hspace{20mm} (8) 
\end{equation}
According to (7) (8), using the vertical coordinate transformation system under non-hydrostatic equilibrium conditions [12], the equation group (1-6) can be transformed into a P coordinate equation system based on the broad static balance:
\begin{equation}
\nonumber
\hspace{0mm}
\frac{d \, V}{dt}
\: + f \: K \times  V 
\; = \;
 - \: \frac{T}{T^{\ast}} \: \nabla \Phi
 \: + \: E_a
\: ,
\hspace{20mm} (9) 
\end{equation}
\begin{equation}
\nonumber
\hspace{0mm}
\frac{d \, \Phi}{dP}
\; = \;
 - \: \frac{R_v \: T^{\ast}}{P} 
\: ,
\hspace{20mm} (10) 
\end{equation}
\begin{equation}
\nonumber
\hspace{0mm}
\frac{d \, \ln(T^{\ast}/T)}{dt}
\: + \:
\frac{\partial u}{\partial x} 
\: + \:
\frac{\partial v}{\partial y} 
\: + \:
\frac{\partial \omega}{\partial P} 
\; = \;
 0
\: ,
\hspace{20mm} (11) 
\end{equation}
\begin{equation}
\nonumber
\hspace{0mm}
\frac{d \, T^{\ast}}{dt}
\: - \:
 \frac{R_v}{C_p} 
 \: \frac{T^{\ast}}{P}
 \; \omega
\; = \;
\frac{1}{C_p} \: \frac{\delta Q}{\delta t} 
\: ,
\hspace{20mm} (12) 
\end{equation}
\begin{equation}
\nonumber
\hspace{0mm}
\frac{d^2 \, \Phi}{dt^2}
\; = \;
 \left( \frac{T}{T^{\ast}} \: - \: 1 \right)
\: g^2
\: .
\hspace{20mm} (13) 
\end{equation}

From (9-13), we can get the following vertical vortex (hereinafter referred to as the position vortex):
\begin{equation}
\nonumber
\hspace{0mm}
\frac{dS}{dt}
\; = \;
S \left( \:
 A
\: + \:
 B
\: + \:
 C
\: + \:
 D
\: + \:
 E
\: + \:
 F
\: + \:
 G
\: \right)
\: ,
\hspace{20mm} (14) 
\end{equation}
where
\begin{equation}
\nonumber
\hspace{0mm}
A \; = \; {\left( \frac{\partial \theta^{\ast}}{\partial P} \right)}^{-1}
\frac{\partial }{\partial P}
\: \left[ \:
\frac{(P/P_0)^{-R_d/C_p}}{C_p}
\:
\frac{\delta Q}{\delta t}
\: \right]
\: ,
\hspace{20mm} (15) 
\end{equation}
\begin{equation}
\nonumber
\hspace{0mm}
B \; = \; 
f^{-1} \:
{\left( \frac{\partial \theta^{\ast}}{\partial P} \right)}^{-1}
\: \left( \:
   - \: \frac{\partial \Phi}{\partial y}
        \frac{\partial \theta^{\ast}}{\partial x}
\: + \: \frac{\partial \Phi}{\partial x}
        \frac{\partial \theta^{\ast}}{\partial y}
\: \right)
\; \:
\frac{\partial}{\partial P}
  \left(
    \frac{\theta}{\theta^{\ast}}
  \right) 
\: ,
\hspace{20mm} (16) 
\end{equation}
\begin{equation}
\nonumber
\hspace{0mm}
C \; = \; 
{\left( f + \zeta \right)}^{-1} \:
\: \left[ \:
        \frac{\partial \Phi}{\partial x} \:
           \frac{\partial}{\partial y}\!
              \left(
           \frac{\theta}{\theta^{\ast}}
              \right)
\: - \:
        \frac{\partial \Phi}{\partial y} \:
           \frac{\partial}{\partial x}\!
              \left(
           \frac{\theta}{\theta^{\ast}}
              \right)
\: \right]
\: ,
\hspace{20mm} (17) 
\end{equation}
\begin{equation}
\nonumber
\hspace{0mm}
D \; = \; 
f^{-1} \:
{\left( \frac{\partial \theta^{\ast}}{\partial P} \right)}^{-1}
\: \left( \:
        \frac{\partial E_y}{\partial y} \;\;
        \frac{\partial \theta^{\ast}}{\partial x}
\: - \: \frac{\partial E_x}{\partial x} \;\;
        \frac{\partial \theta^{\ast}}{\partial y}
\: \right)
\: ,
\hspace{20mm} (18) 
\end{equation}
\begin{equation}
\nonumber
\hspace{0mm}
E \; = \; 
{\left( f + \zeta \right)}^{-1} \:
\: \left( \:
        \frac{\partial E_y}{\partial x} 
\: - \:
        \frac{\partial E_x}{\partial y}
\: \right)
\: ,
\hspace{20mm} (19) 
\end{equation}
\begin{equation}
\nonumber
\hspace{0mm}
F \; = \; 
f^{-1} \:
{\left( \frac{\partial \theta^{\ast}}{\partial P} \right)}^{-1}
\: \left[ \:
        \frac{\partial \theta^{\ast}}{\partial y} \;
        \frac{\partial}{\partial P}\!
        \left( - \: \frac{d u}{dt} \right)
\: - \: \frac{\partial \theta^{\ast}}{\partial x} \;
         \frac{\partial}{\partial P}\!
        \left( - \: \frac{d v}{dt} \right)
\: \right]
\: ,
\hspace{15mm} (20) 
\end{equation}
\begin{equation}
\nonumber
\hspace{0mm}
G \; = \; 
{\left( f + \zeta \right)}^{-1} \:
\: \left( \:
        \frac{\partial u}{\partial P} \:
           \frac{\partial \omega}{\partial y}
\: - \:
        \frac{\partial v}{\partial P} \:
        \frac{\partial \omega}{\partial x}
\: \right)
\: ,
\hspace{20mm} (21) 
\end{equation}
\begin{equation}
\nonumber
\hspace{0mm}
G \; = \; 
\frac{\theta^{\ast}}{\theta} \;
\left(
\frac{\partial \theta^{\ast}}{\partial P} 
\right) \;
{\left( f + \zeta \right)}
\: ,
\hspace{20mm} (22) 
\end{equation}
where $\theta^{\ast} = T^{\ast} \: (P/P_0)^{-R_d/C_p}$ 
is the generalized potential temperature, 
$\theta=T \: (P/P_0)^{-R_d/C_p}$ is the potential temperature, 
$S$ is the vertical wet vortex (hereinafter referred to as the wet vortex).

\begin{itemize}
\item
The term $A$ is the contribution of the non-adiabatic heating except the latent heat of condensation to the change of the vertices by changing the vertical stratification.
\item
Item $B$ is a contribution to the vortex change associated with the condensed latent heat associated with the vertical unevenness of the static imbalance caused by condensation latent heat.
\item
The $C$ term is a contribution of the condensed latent heat to the vortex change associated with the level unevenness of the static imbalance caused by the latent heat of condensation.
\item
The $D$ term is the contribution of the external force field to the change of the vortex by changing the stratification.
\item
The $E$ term is the contribution of the external force field to the change of the vortex by changing the wind field.
Since the external force field that affects the atmospheric movement is mostly a potential field such as the astronomical tide field, this is always zero.
\item
The $F$ term is the contribution of the wind field development by changing the vortex change of the stratification.
\item
The $G$ term is the contribution of the vortex tube effect to the potential vortex.
\end{itemize}

It is important to note that the size of these items depends to a large extent on the denominator factors 
$\partial \theta^{\ast} / \partial P \: (A, B, D, F \; \mbox{terms})$, 
$f (B, D, F \; \mbox{terms})$, and 
$(f+ \xi) (C, E, G \; \mbox{terms})$.

Equation (14) is the basic equation we discuss.

\vspace*{4mm}
{\bf \large 2)  \underline{Potential vorticity}}
\vspace*{2mm}
\label{2_Potential_orticity}

From equation (14), it can be seen that the vortices are always non-conservative as long as there are non-adiabatic heating and external force fields on the right hand.
However, if the contribution of the vortex change on the right side of equation (14) is less than or equal to the vortex level and vertical feed, then the vortex is considered to be either conservative or quasi-conservative.
Therefore, if E=0 is noted, then according to the dimensionless form of (14):
\begin{align}
\nonumber
& \hspace{0mm}
\varepsilon \: 
\frac{\partial S_1}{\partial t_1} 
\: + \: R \: 
   \left(
       u_1 \: \frac{\partial S_1}{\partial x_1}
       \: + \:
       v_1 \: \frac{\partial S_1}{\partial y_1}      
   \right)
\: + \: 
\frac{W}{f \: H_P} \;
   \left(
       \omega_1 \: \frac{\partial S_1}{\partial P_1}
   \right)
\; = \; 
\\
\nonumber
& \hspace{60mm}
S_1 \:
O \! \left[ \: 
f^{-1} \:
\left( \:
 A
\: + \:
 B
\: + \:
 C
\: + \:
 D
\: + \:
 F
\: + \:
 G
\: \right)
\: \right]
\
\: .
\hspace{5mm} (23) 
\end{align}
The vorticity non-conservation criterion can be obtained as:
\begin{equation}
\nonumber
O \! \left[ \: 
f^{-1} \:
\left( \:
 A
\: + \:
 B
\: + \:
 C
\: + \:
 D
\: + \:
 F
\: + \:
 G
\: \right)
\: \right]
\; \geq \;
\max\left[ \:
 R \: \frac{W}{f \: H_P}
\: \right]
\: .
\hspace{10mm} (24) 
\end{equation}
For mid-latitude mid-scale and large-scale atmospheric movements, there are generally:
\begin{align}
\nonumber
 \Delta_h \left( \theta / \theta^{\ast} \right)
 & < \; 10^{-1} 
\: , 
\hspace{10mm} (25) 
 \mbox{\phantom{$\displaystyle \frac{W}{f \: H_P}$}}
\\
\nonumber
 \Delta_l \left( \theta / \theta^{\ast} \right)
 & < \; 10^{-1}
\: , 
\hspace{10mm} (26) 
 \mbox{\phantom{$\displaystyle \frac{W}{f \: H_P}$}}
\\
\nonumber
 R \; \sim \; \frac{W}{f \: H_P}
 & \leq \; 10^{0}
\: .
\hspace{10mm} (27) 
\end{align}
The subscript of symbol $\Delta$ are ``$l$'' for the horizontal direction and ``$h$'' for the vertical direction.

It is not difficult to prove that
\begin{align}
\nonumber
 O\left( \: f^{-1} \: C \: \right)
 & < \; R
\: , \hspace{10mm} (28)
 \mbox{\phantom{$\displaystyle \frac{W}{f \: H_P}$}}
\\
\nonumber
 O\left( \: f^{-1} \: G \: \right)
 & \leq \; R
\: .
\hspace{10mm} (29) 
\end{align}
That is to say, at this time, the vorticity non-conservation criterion (24) can be simplified as:
\begin{equation}
\nonumber
O \! \left[ \: 
f^{-1} \:
\left( \:
 A
\: + \:
 B
\: + \:
 D
\: + \:
 F
\: \right)
\: \right]
\; \geq \;
\max\left[ \:
 R \: \frac{W}{f \: H_P}
\: \right]
\: .
\hspace{10mm} (30) 
\end{equation}
(namely with $C$ and $G$ removed))

\vspace*{4mm}
{\bf \large 3)  \underline{Characteristics of non-conservative process of the wet vortex}}
\vspace*{2mm}
\label{3_Char_PV}

According to the above criterion (30), it can be seen that for the non-conservation process of the vortex (marked by the subscript N), there should be
\begin{equation}
\nonumber
O \! \left( \:  \varepsilon_N \: \right)
\; = \;
O \! \left[ \: 
f^{-1} \:
\left( \:
 A
\: + \:
 B
\: + \:
 C
\: + \:
 D
\: + \:
 E
\: + \:
 F
\: \right)
\: \right]
\; \geq \;
O \! \left( \:  R_N \: \right)
\: .
\hspace{10mm} (31) 
\end{equation}
For the vortex conservation and conservation motion (with the subscript C as the mark to distinguish the non-conservative motion), there are generally:
\begin{equation}
\nonumber
O \! \left( \:  \varepsilon_C \: \right)
\; < \;
O \! \left( \:  R_C \: \right)
\: .
\hspace{10mm} (32) 
\end{equation}
Therefore, for the same scale $R_N \: \sim \: R_c$, there are:
\begin{equation}
\nonumber
O \! \left( \:  \varepsilon_N \: \right)
\; > \;
O \! \left( \:  \varepsilon_C \: \right)
\: .
\hspace{10mm} (33) 
\end{equation}
This shows that the time scale $\tau_N$ of the non-conservative motion of the vortex is much smaller than the time scale $\tau_C$ in the case of conservation, for example, for mesoscale atmospheric motion 
$R \: \sim \: 10^0$,
$\tau_C \: \sim \: 10^4$, and
$\tau_N \: < \: 10^4$ seconds,  
and for Large-scale movement 
$R \: \sim \: 10^{-1}$ and
$\tau_C \: \sim \: 10^5$ seconds.

However, $\tau_C \: \sim \: 10^5$ seconds is one of the characteristics of the vorticity non-conservation process: it is a fast process relative to the vortex conservation of atmospheric motion.

If it can be proved that the real circulation adjustment is closely related to the vorticity non-conservation process (we will discuss this problem in another article), then this feature can fully explain the fact that the rapid circulation of the atmospheric circulation and the relative stability change occur.

In addition, from the vorticity non-conservation criterion (24) and the $A$, $B$, $C$, $D$, $F$ and $G$ given by the expressions (15), (16), (17), (18), (20) and (21), it can be seen that for free mesoscale and large-scale atmospheric motion due to the non-adiabatic heating 
$\delta Q / \delta t$ 
other than condensation release, the force field E and 
$\Delta ( \theta , \theta^{\ast} )$, $\Delta_h = | (dV/dt) |$, 
the value of the scroll item is relatively small, even if there is disturbance.

The criterion (24) $\Delta$ is generally difficult to satisfy.
However, for wet neutral gas, low latitude area or anticyclonic area, the wet vorticity will become important due to the role of the denominator 
$\partial \theta^{\ast} / \partial P$, $f$ and $(f+\xi)$.

For the mid-latitudes, the sufficient condition for the occurrence of the vorticity non-conservation process is that the dynamic neutralization tends to occur in the wet neutral layer and the vortex vortex.

And according to (18) (24), and using the tidal potential expression (see [13]), even a small amount of astronomical tidal force field ($E \sim 10^{-6}$~g), when the other external sources are fundamentally balanced, satisfies:
\begin{equation}
\nonumber
O \! 
\left( \:
 A
\: + \:
 B
\: + \:
 C
\: + \:
 F
\: + \:
 G
\: \right)
\; \leq \;
O \! \left( \:  D \: \right)
\: .
\hspace{10mm} (34) 
\end{equation}
For the case of $\partial \theta^{\ast} / \partial P$ tending to be zero, the sufficient condition is:
\begin{equation}
\nonumber
O \! \left( \:
\frac{\partial \theta^{\ast} / \partial x}
     {\partial \theta^{\ast} / \partial z}
  \: \right)
\; > \;
O \! \left( \:  
\frac{a \: f^2}
     {2  \: E}
  \; R
  \: \right)
\: ,
\hspace{10mm} (35) 
\end{equation}
where $a$ is the radius of the Earth.
The change in the tidal force field can also cause the vorticity to follow the non-conservation process.

\vspace*{4mm}
{\bf \large 4)  \underline{Dynamic characteristics of wet vorticity without constant motion}}
\vspace*{2mm}
\label{4_Char_PV}

Here we only discuss the mid-latitude and mid-latitude wet vorticity non-conservation processes in the mid-latitude, and the vorticity in the low-latitude region is similar to the non-conservation process.

According to the above analysis, for the non-conservation process of large-scale and mesoscale vortices, the advection effect can be omitted.

The static imbalance level caused by the uneven condensation of water vapor level, the vortex effect, and therefore the vorticity equation, the Degree equation, the lumped parameter equation, and the continuous equation can be simplified to:

\hspace{20mm} Eqs.~(36) - (39)

And according to Eqs~(36)-(39):

\hspace{20mm} Eqs.~(40) - (41)

The necessary and sufficient condition for the non-conservation process of the vortex in the mid-latitude is that $\partial \theta^{\ast} / \partial P$ tends to zero.
The dynamic undulation of the effects of the vorticity is affected. 
Therefore, as an approximation, for the mid-latitude vorticity indefinite process, $\partial \theta^{\ast} / \partial P \approx 0$ can be assumed. 

Then Eqs~(40)-(41) can be simplified as:

\hspace{20mm} Eqs.~(42) - (43)

The equations (43) and (47) show that, in the non-conservative process of the vortex, the non-adiabatic heating, the vertical unevenness of the water vapor condensation and the external force field change the divergence field by the forced field to produce the ground deviation.

Therefore, even if the initial field is leveled to the balance field, it will also cause the excitement of the ground rotation.

\vspace*{4mm}
{\bf \large 5)  \underline{Possible physical mechanism for changes in weather}
\vspace*{1mm} \\ \hspace*{9mm}
\underline{Adaptation, development and stimulation process}}
\vspace*{2mm}
\label{5_Char_PV}

From the existing results (see [14]), it can be seen that when a high-level non-earth-turning wind field emerges in a certain area, it will immediately lead to the evolution of the circulation.

The initial changes are very fast. 
After a short period of time, the atmospheric flow field will be close to the quasi-ground balance. 
This period of time is called the geostrophic {\bf adaptation process}.

In these two processes, their potential vorticity is always quasi-conservative.

However, according to the above discussion, in the {\bf development process} in the wet neutral layer ($\partial \theta^{\ast} / \partial P > 0$) region, if non-adiabatic heating, water vapor condensation latent heat release, external force field and wind field development on the layered feedback dynamic balance due to some causes are destroyed, so that the vorticity non-conservation criterion (30) can be satisfied.

Then, in those areas, non-adiabatic heating, condensation of latent heat release, external force field and wind field development feedback will affect the dynamic imbalance. 
This leads to the non-conservative process of the vortex, and the process is a fast process relative to the quasi-conservative process. 
The physical essence of this process is non-adiabatic heating, vertical non-uniformity of water condensation and dynamic field of external force. 
Balance the forced change of the stratification and ground rotation deviation (in the quasi-ground transition initial field, that is, the {\bf excitation} of the ground rotation deviation)
For this process, it is called the {\bf excitation process}. 

During the {\bf excitation process}, since $\partial \theta^{\ast} / \partial P$ tends to zero, it can be proved that the ground rotation deviation is very slow to spread around the gravity wave dispersion, so it is mainly through the change and influence of the layer junction. 
The effects of the vorticity changes are re-adjusted and {\bf adapted} to non-adiabatic heating, vertical condensation of water vapor condensation and external force fields (relative to atmospheric pressure gradient forces in the atmospheric system, in terms of Coriolis force, they are foreign sources).
The dynamic imbalance of the field tends to rebalance, so that the potential vortex reaches the conservation again.
Therefore, the process of stimulating can also be called the external {\bf adaptation process}.

After the exogenous {\bf adaptation process} is completed, since the $\partial \theta^{\ast} / \partial P$ is no longer equal to zero, the geostrophic deviation generated by the agitation process will be accelerated by the gravity wave to the four-week dispersion, that is, into a new round of {\bf adaptation} and {\bf development} processes.

The possible physical mechanism of change in weather is the process of this process, {\bf adapting} to the process and {\bf development} process, and then {\bf initiating} the cycle of the process.
The contact mechanism between the astronomical tide field and the atmospheric movement will be discussed separately.

\vspace*{4mm}
{\bf \large \underline{References}}
\vspace*{2mm}
\label{References}

[1] 
Xie Yibing.
Astronomical Dynamics of Wet and Oblique Air.
Rainstorm Collection, 
Jilin People's Publishing House, 1978, 1-15.

[2]
Wang Liangming.
Basic equations and main features of saturated aerodynamics.
Journal of Gas Imaging, 1980, 38(1), 44-50.

[3]
Hoskins B.J, McIntyre M.E. and Robertson A.W.
On the use and significance of isentropic potential 
vorticity maps.
Quart. J. Roy. Meteor. Soc.,
1985, 111:877-946.

[4] Bennetts D.A. and Hoskins B.J.
Conditional symmetric instability - a possible
explanation for frontal rainbands,
Quart. J. Roy. Meteor. Soc.,
1979, 105:945-962.

[5]
Wu Guoxiong.
Wet Potential Vortex and Inclined Vortex Development.
Meteorological Report, 195, 53(4), 387-405.

[6]
Zhu Gangen et al.
Principles and Methods of Astronomy.
Journalism Publishing House, 1981, 228.

[7]
Zhu Zhenquan.
Lunar Phase and Cold Air Activity Preview.
Gas Science, 1982, 6(1), 44-51.

[8]
Ren Zhenqiu.
The tidal force of the sun and the typhoon intensity change.
Meteorology, 1975, (9): 18-20.

[9]
Wang Xingrong. 
Relationship between the astronomical tides and the 
occurrence of cold tides under different geomorphological 
conditions. 
Meteorological Journal, 1990, 48(2) 239-241.

[10]
Wang Xingrong, Yan Xuefeng, Mei Yu. 
Relationship between subtropical high pressure 
activity and near-month lunar phase. 
Gas Science, 1995, 19(5), 636-640.

[11] 
Wang Xingrong, Shi Zhenling et al. 
The relationship between the high pressure 
activity of the subtropical zone and the operation 
of the sun and the moon. 
Hot Belt Gas Imaging, 1997, 13(1): 92-96.

[12]
Wang Xingrong et al. 
Discussion on several issues in wet air dynamics. 
Gas Science, 1995, 15(1).

[13]
Lamb H.
Hydrodynamics. 
Cambridge univ. 1932, 359.

[14] 
Zeng Qingcun. 
Mathematical and Physical Basis of Numerical 
Weather Report (Vol.1).
Science Press, 1979, 202-494.

          \newpage 

\setcounter{equation}{0} 

\begin{center} 
{\Large 
{\bf Appendix~3:~}{\it A (crude and partial) translation  of the Chinese paper\/}\\
\vspace{2mm}
{\bf Wang}${}^{(1)}$ Xingrong, {\bf Wu}${}^{(1)}$ Kejun and {\bf Shi}${}^{(1)}$ Chun'e (1999).\\
{\bf 
\vspace{1mm}
{\color{blue}
Introduction of condensation probability function 
and non-uniform saturated wet air dynamics equations.
}\\
\vspace{2mm}
Journal of Tropical Meteorology, 1999, 15 (1), 64-70.
}
}
\end{center}
\vspace{0 mm}

(1) Anhui Institute of Meteorological Sciences, Anhui Hefei

\vspace{2 mm}

{\noindent \small WWW links:
\url{http://www.itmm.org.cn/rdqxxb/ch/reader/view_abstract.aspx?file_no=19990108&flag=1}
}

{\noindent \small The PDF: $\;$
\url{http://www.itmm.gov.cn/rdqxxb/ch/reader/create_pdf.aspx?file_no=19990108}
}

\vspace*{4mm}
{\bf \large \underline{Abstract}}
\vspace*{2mm}
\label{Summary}

Introducing the clotting probability function and abandoning the assumption that the condensing process begins at a critical relative value, so that the wet aerodynamics study is free from the irrational requirement that the whole atmosphere is always saturated, and the non-uniform saturated wet air dynamics is established. 
Equations, and some simplified forms of equations are given under certain assumptions.

\vspace*{4mm}
{\bf \large 1)  \underline{Introduction}}
\vspace*{2mm}
\label{1_Intro}

Water vapor is closely related to weather activities. With the deepening of typhoon and storm analysis and forecasting, people are paying more and more attention to the role of water vapor in the occurrence and development of weather systems [1-3].

Wang Xingrong et al. also used a rigorous mathematical derivation to obtain a vertical coordinate transformation formula that is not based on static balance, and used it to establish a P-coordinate saturated wet aerodynamics based on strict mathematical proof based on static balance Equations [4].

However, all of the above studies have a significant defect, that is, the whole atmosphere is always saturated at all times, and in the actual large-scale motion, there is a downdraft in the ascending airflow, and the initial saturated humid air in the sinking motion zone also changes. Is not saturated.
That is to say, the actual wet air is usually non-uniformly saturated (that is, partially saturated in some places).
Therefore, whether these theories describe the actual wet air problem is reasonable and the scope of application is worthy of scrutiny.
Analysis of the causes of this defect, it is not difficult to find, mainly the current understanding and treatment of the condensation process of dynamic meteorology.

In the current dynamic meteorological and numerical prediction work, the important role of water vapor is usually considered by introducing condensation heating and water vapor equations.

One important assumption is that the condensation process begins at a critical relative value, which forces the introduction of a discontinuous $\delta$ function in the water vapor equation.

Although the discontinuity of the $\delta$ function does not make the numerical prediction differential calculation difficult, it makes the dynamic analysis of the wet aerodynamics at the boundary between the condensation zone and the non-condensation zone extremely difficult, forcing various theoretical studies of wet aerodynamics assuming that the entire atmosphere is always saturated.

In view of the reasons for this defect, this paper understands the process of treating the condensation of water vapor from another angle, so that the wet aerodynamics study can get rid of the unreasonable requirement of maintaining saturation everywhere, and establish a non-uniform saturated wet air dynamics equation.

\vspace*{4mm}
{\bf \large 1)  \underline{Introduction of the condensation probability function}}
\vspace*{2mm}
\label{2_cond_prop_func}

It is well known that the atmosphere contains a large number of condensation nuclei which have a hygroscopic function similar to that of defatted hair.

Therefore, according to the theory of molecular statistics, it can be assumed that under certain atmospheric conditions (pressure $P$, temperature $T$, and hygroscopic condensed hygroscopic concentration $\rho \: N$), the ratio of the infinite small unit to the wet $q'$, not necessarily equal to $q$, but a statistical distribution satisfying the probability function $f ( P, T , q, \rho \: N, q' / q_s )$, the probability function is satisfied
\begin{equation}
\nonumber
\hspace{40mm}
\int_0^\infty
f \! \left(
P, T, q, \rho \: N, q'/q_s
\right)
\:
d \!
\left(
q'/q_s
\right)
\; = \;
1
\: ,
\hspace{30mm} (1) 
\end{equation}
\begin{equation}
\nonumber
\hspace{40mm}
\int_0^\infty
q' \:
f \! \left(
P, T, q, \rho \: N, q'/q_s
\right)
\:
d \!
\left(
q'/q_s
\right)
\; = \;
q
\: .
\hspace{30mm} (2) 
\end{equation}

For units with a moisture $q'$ greater than $q_s$, due to the action of hygroscopic condensation nuclei, there is always a certain amount of liquid water, and the water vapor of this unit is in the dynamic equilibrium of evaporation and condensation.

According to this assumption, for an atmosphere containing a certain amount of water vapor, a part of the water vapor is involved in the dynamic equilibrium of condensation and evaporation with liquid water, and a certain amount of liquid water is maintained in the atmosphere. The minimum content of liquid water per unit volume $m$ can generally be written as
\begin{equation}
\nonumber
\hspace{10mm}
m \: = \:
q_s \:
\int_1^\infty
\left(
q'/q_s
\right)
\:
G \! \left(
P, T, \rho \: N, q'/q_s
\right)
\:
f \! \left(
P, T, q, \rho \: N, q'/q_s
\right)
\:
d \!
\left(
q'/q_s
\right)
\: ,
\hspace{20mm} (3) 
\end{equation}
where $q' \: G ( P, T , \rho \:N, q'/q_s )$ is the smallest possible liquid water content contained in an infinitesimal unit of wet $q'$ ($q' > q_s$).

Equation (3) can be rewritten as
\begin{equation}
\nonumber
\hspace{10mm}
m \: = \:
\frac{q_s}{2} \:
\int_1^\infty
G \! \left(
P, T, \rho \: N, q'/q_s
\right)
\:
f \! \left(
P, T, q, \rho \: N, q'/q_s
\right)
\:
d
{\left(
q'/q_s
\right)}^2
\: .
\hspace{20mm} (4) 
\end{equation}

In general, when the air pressure $P$, the temperature $T$, and the condensation hygroscopic concentration $\rho \: N$ are the same, the larger $q$ is, the larger $m$ is.
Set the relationship between the two
\begin{equation}
\nonumber
\hspace{10mm}
q \; = \;
h \left(
P, T, \rho \: N, q
\right)
\; m
\: ,
\hspace{20mm} (5) 
\end{equation}
where $h ( P , T , \rho \: N , q )$ is the specific wetness of the unit liquid water content.
From (4) and (5):
\begin{equation}
\nonumber
\hspace{10mm}
\frac{dq}{dt} 
\; = \;
\frac{d
\left[ \:
k \left(
P, T, q, \rho \: N
\right)
\: q_s
\: \right]
}{dt} 
\: ,
\hspace{20mm} (6) 
\end{equation}
where
\begin{equation}
\nonumber
\hspace{0mm}
k \left(
P, T, \rho \: N, q
\right)
\: = \:
\frac{
h \! \left(P, T, \rho \: N, q \right)
}{2} \:
\int_1^\infty
G \! \left(
P, T, \rho \: N, q'/q_s
\right)
\:
f \! \left(
P, T, q, \rho \: N, q'/q_s
\right)
\:
d
{\left(
q'/q_s
\right)}^2
\: ,
\hspace{5mm} (7) 
\end{equation}
We define this as a function of condensation probability.

According to the law of conservation of water vapor and liquid water, there is
\begin{equation}
\nonumber
\hspace{10mm}
   - \: \frac{dq}{dt} 
\; = \;
   - \: \frac{dm}{dt} 
\: + \: \eta 
\: ,
\hspace{20mm} (8) 
\end{equation}
where $\eta$ is the liquid water content exceeding $m$, and if all settles, the sedimentation rate of liquid water.

As everyone knows, in the case of saturation
\begin{equation}
\nonumber
\hspace{10mm}
\frac{dq}{dt} 
\; = \;
\frac{dq_s}{dt} 
\: ,
\hspace{20mm} (9) 
\end{equation}
and in the case of complete drying
\begin{equation}
\nonumber
\hspace{10mm}
\frac{dq}{dt} 
\; = \;
0 
\: .
\hspace{20mm} (10) 
\end{equation}
From (6), (9), (10), it is not difficult to derive
\begin{equation}
\nonumber
\hspace{0mm}
k \left(P, T, 0, \rho \: N \right)
\; = \;
0
\: ,
\hspace{20mm} (11) 
\end{equation}
\vspace{-4mm}
\begin{equation}
\nonumber
\hspace{0mm}
k \left(P, T, q_s, \rho \: N \right)
\; = \;
1
\: .
\hspace{20mm} (12) 
\end{equation}

According to the characteristics of the condensation probability function $k ( P , T , q, \rho \: N )$ reflected by (11) and (12), the function form of $k ( P , T , q, \rho \: N )$ can be rewritten as
\begin{equation}
\nonumber
\hspace{0mm}
k \left(P, T, q, \rho \: N \right)
\; = \;
{\left(
q/q_s
\right)}^{\alpha}
\: ,
\hspace{20mm} (13) 
\end{equation}
where $\alpha$ is a positive real function of $P , T , q, \rho \: N$.

If the functional form of $m$, $\alpha$ is known, combined with the equation for controlling the concentration of condensed hygroscopic particles, then
\begin{equation}
\nonumber
\hspace{0mm}
\frac{d \, \rho \, N}{dt}
\; = \;
\nabla \: . \: 
\left(
\sigma \: . \: \nabla \, \rho \, N
\right)
\: - \: \Lambda \; \rho \, N
\: + \: V_g 
     \: \frac{\partial \, \rho \, N}{\partial z}
\: ,
\hspace{20mm} (14) 
\end{equation}
where $\sigma$ is the turbulent diffusion coefficient vector, $\Lambda$ is the precipitation cleaning coefficient, and $V_g$ is the gravity deposition average velocity.

According to the above physical image mode, using the existing dynamic work, a more rigorous frictionless, wet adiabatic, non-uniform saturated wet aerodynamic equation can be obtained:
\begin{equation}
\nonumber
\hspace{0mm}
\frac{d \, V}{dt}
\: + f \: K \times  V 
\; = \;
 - \: \frac{1}{\rho} \: \nabla P
\: ,
\hspace{20mm} (15) 
\end{equation}
\begin{equation}
\nonumber
\hspace{0mm}
\frac{d \, w}{dt}
\; = \;
 - \: \frac{1}{\rho} 
   \: \frac{\partial \, P}{\partial z}
\: - \: g
\: ,
\hspace{20mm} (16) 
\end{equation}
\begin{equation}
\nonumber
\hspace{0mm}
\frac{d \, T}{dt}
\: - \: \frac{1}{C_p} \: \frac{1}{\rho} 
   \: \frac{d \, P}{d t}
\; = \;
 - \: \frac{L}{C_p} \;
 \frac{
 d \left[ \:
 {(q/q_s)}^{\alpha} \: q_s
 \: \right]
 }{dt} 
\: ,
\hspace{20mm} (17) 
\end{equation}
\begin{equation}
\nonumber
\hspace{0mm}
P
\; = \;
 \rho \: R \: T
\: ,
\hspace{20mm} (18) 
\end{equation}
\begin{equation}
\nonumber
\hspace{0mm}
\frac{1}{\rho} \:
\frac{d \, \rho}{dt}
\: + \:
\nabla \, . \, V
\: + \:
\frac{\partial \, \omega}{\partial z}
\; = \;
 0
\: ,
\hspace{20mm} (19) 
\end{equation}
\begin{equation}
\nonumber
\hspace{0mm}
\frac{d \, q}{dt}
\; = \;
 \frac{
 d \left[ \:
 {(q/q_s)}^{\alpha} \: q_s
 \: \right]
 }{dt} 
\: ,
\hspace{20mm} (20) 
\end{equation}
\begin{equation}
\nonumber
\hspace{0mm}
- \:
 \frac{
 d \left[ \:
 {(q/q_s)}^{\alpha} \: q_s
 \: \right]
 }{dt} 
\; = \;
\frac{d \, m}{dt}
\: + \: \eta
\: ,
\hspace{20mm} (21) 
\end{equation}
\begin{equation}
\nonumber
\hspace{0mm}
\frac{d \, \rho \, N}{dt}
\; = \;
\nabla \: . \: 
\left(
\sigma \: . \: \nabla \, \rho \, N
\right)
\: - \: \Lambda \; \rho \, N
\: + \: V_g 
     \: \frac{\partial \, \rho \, N}{\partial z}
\: ,
\hspace{20mm} (22) 
\end{equation}
\begin{equation}
\nonumber
\hspace{0mm}
\alpha
\; = \;
f_1  \left(P, T, q, \rho \: N \right)
\: ,
\hspace{20mm} (23) 
\end{equation}
\begin{equation}
\nonumber
\hspace{0mm}
m
\; = \;
f_2  \left(P, T, q, \rho \: N \right)
\: ,
\hspace{20mm} (24) 
\end{equation}
\begin{equation}
\nonumber
\hspace{0mm}
q_s
\; = \;
f_3  \left(P, T \right)
\: .
\hspace{20mm} (25) 
\end{equation}

Obviously, this system of equations is suitable for atmospheric motion with various humidity states.
In the ideal state, when $q$ is $0$ and $q_s$ respectively, it degenerates into a dry atmospheric dynamics equation and a saturated wet aerodynamic equation.

Moreover, since this system of equations is not formed on the basis of the irrational assumption that the ``condensation process begins at a critical relative value'', there is no difficulties in the kinetic analysis of the boundary of the condensation zone, due to the continuous success of the $\delta$ function.

However, although the physical mode is more rigorous without considering the effects of condensation, and due to the lack of a basic understanding of $f_1$, $f_2$, it is very difficult to strictly follow this equation to deal with practical problems.
However, if you simplify the equations for some specific problems, you can solve these problems by bypassing the discussion of $f_1$ and $f_2$.

\vspace*{4mm}
{\bf \large 3) \underline{Some 
simplified equations for non-uniform }
\vspace{2mm} \\ \hspace*{9mm}
\underline{saturated wet air dynamics}
}
\vspace*{2mm}
\label{3_saturated}
\vspace*{2mm}

\vspace*{4mm}
{\bf \large \hspace{10mm} 3.1)  \underline{Non-uniform 
saturated wet aerodynamic.
}}
\vspace*{2mm}
\label{31_saturated}

As the most simplified form, you can assume that $\alpha$ is a constant, so that equations (15)-(25) can be simplified to
\begin{equation}
\nonumber
\hspace{0mm}
\frac{d \, V}{dt}
\: + f \: K \times  V 
\; = \;
 - \: \frac{1}{\rho} \: \nabla P
\: ,
\hspace{20mm} (26) 
\end{equation}
\begin{equation}
\nonumber
\hspace{0mm}
\frac{d \, w}{dt}
\; = \;
 - \: \frac{1}{\rho} 
   \: \frac{\partial \, P}{\partial z}
\: - \: g
\: ,
\hspace{20mm} (27) 
\end{equation}
\begin{equation}
\nonumber
\hspace{0mm}
\frac{d \, T}{dt}
\: - \: \frac{1}{C_p} \: \frac{1}{\rho} 
   \: \frac{d \, P}{d t}
\; = \;
 - \: \frac{L}{C_p} \;
 \frac{
 d \left[ \:
 {(q/q_s)}^{\alpha} \: q_s
 \: \right]
 }{dt} 
\: ,
\hspace{20mm} (28) 
\end{equation}
\begin{equation}
\nonumber
\hspace{0mm}
P
\; = \;
 \rho \: R \: T
\: ,
\hspace{20mm} (29) 
\end{equation}
\begin{equation}
\nonumber
\hspace{0mm}
\frac{1}{\rho} \:
\frac{d \, \rho}{dt}
\: + \:
\nabla \, . \, V
\: + \:
\frac{\partial \, \omega}{\partial z}
\; = \;
 0
\: ,
\hspace{20mm} (30) 
\end{equation}
\begin{equation}
\nonumber
\hspace{0mm}
\frac{d \, q}{dt}
\; = \;
 \frac{
 d \left[ \:
 {(q/q_s)}^{\alpha} \: q_s
 \: \right]
 }{dt} 
\: ,
\hspace{20mm} (31) 
\end{equation}

If we follow the literature [2] and [4], we define the generalized temperature of non-uniform saturated wet air as
\begin{equation}
\nonumber
\hspace{0mm}
T^{\ast \ast}
\; = \;
T \;
 \exp
 \left[ \:
 \frac{L}{C_p} \:
 \frac{q_s}{T} \:
 {\left( \frac{q}{q_s} \right)}^{\alpha}
 \: \right]
\: ,
\hspace{20mm} (32) 
\end{equation}
According to the derivation process of the literature [4], the $P$-coordinate system non-uniform saturated wet aerodynamics equations with the same form as the P-coordinate system of saturated aerodynamic equations in [4] can be obtained:
\begin{equation}
\nonumber
\hspace{0mm}
\frac{d \, u}{dt}
\; = \;
 f \: v
 - \: \frac{T}{T^{\ast \ast}} \: \frac{\partial \Phi}{\partial x}
\: ,
\hspace{20mm} (33) 
\end{equation}
\begin{equation}
\nonumber
\hspace{0mm}
\frac{d \, v}{dt}
\; = \;
 - \: f \: u
 - \: \frac{T}{T^{\ast \ast}} \: \frac{\partial \Phi}{\partial y}
\: ,
\hspace{20mm} (34) 
\end{equation}
\begin{equation}
\nonumber
\hspace{0mm}
\frac{d \, \Phi}{dP}
\; = \;
 - \: \frac{R \: T^{\ast \ast}}{P} 
\: ,
\hspace{20mm} (35) 
\end{equation}
\begin{equation}
\nonumber
\hspace{0mm}
\frac{d \, \ln(T^{\ast \ast}/T)}{dt}
\: + \:
\frac{\partial u}{\partial x} 
\: + \:
\frac{\partial v}{\partial y} 
\: + \:
\frac{\partial \omega}{\partial P} 
\; = \;
 0
\: ,
\hspace{20mm} (36) 
\end{equation}
\begin{equation}
\nonumber
\hspace{0mm}
C_v \: \frac{d \, T^{\ast \ast}}{dt}
\: + \:
 P\: \frac{d \, (R \: T^{\ast \ast} / P)}{dt}
\; = \;
0
\: ,
\hspace{20mm} (37) 
\end{equation}
\begin{equation}
\nonumber
\hspace{0mm}
\frac{d^2 \, \Phi}{dt^2}
\; = \;
 \left( \frac{T}{T^{\ast \ast}} \: - \: 1 \right)
\: g^2
\: .
\hspace{20mm} (38) 
\end{equation}
This indicates that any conclusions derived from the $P$-coordinate saturated wet aerodynamic equations in [4] will apply to the non-uniform saturated humid air obtained from the above assumptions, except for the generalized temperature in non-uniform saturated humid air, where $T^{\ast \ast}$ is replaced in [4] by $T^{\ast}$ given by
\begin{equation}
\nonumber
\hspace{0mm}
T^{\ast}
\; = \;
T \;
 \exp
 \left[ \:
 \frac{L}{C_p} \:
 \frac{q_s}{T} \:
 \: \right]
\: ,
\hspace{20mm} (39) 
\end{equation}
whereas $T^{\ast \ast}$ satisfies (32).

Comparing (32) and (39), it is not difficult to find that $T^{\ast}$ is a special case of $T^{\ast \ast}$ at saturation.

As for the mode and theoretical research, it is better to derive $\alpha$ according to the fact that when the relative humidity $q/q_s = f$ is greater than $0.78$ in the actual atmosphere, the condensation can be found under the action of condensation nuclei [5], if
\begin{equation}
\nonumber
\hspace{0mm}
k
\; = \;
{\left( \:
 \frac{q}{q_s}
 \: \right)}^{\alpha}
\; \leq \; k_0
\: .
\hspace{20mm} (40) 
\end{equation}

As a criterion for not finding condensation, with $k_0$ between $0$ and $1$, it is better to see that it is about $9$ as shown in Table~\ref{Table1}. 
Of course, the value of $k_0$ and the correct determination of $\alpha$ depend on the results of theoretical and numerical simulations with the actual situation.
\begin{table}
\caption{Values of $0.78^\alpha$ and $0.75^\alpha$ under different conditions of $\alpha$.
\label{Table1}
}
\centering
\vspace*{2mm}
\begin{tabular}{|c|ccccccccccc|}
\hline
$\alpha$ & 1 & 2 & 3 & 4 & 5 & 6 & 7 & 8 & 9 & 10 & 11 \\ 
\hline
$0.78^\alpha$ & 0.78 & 0.61 & 0.48 & 0.37 & 0.29 & 0.23 & 0.18 & 0.14 & 0.11 & 0.08 & 0.07 \\ 
\hline
$0.75^\alpha$ & 0.75 & 0.56 & 0.42 & 0.32 & 0.23 & 0.18 & 0.13 & 0.10 & 0.08 & 0.06 & 0.04  \\ 
\hline
\end{tabular}
\vspace*{-4mm} 
\end{table}

It should be noted that if $\alpha$ is infinity, $k$ degenerates into a $\delta$ function.
That is to say, the previous assumption that the ``condensation process starts at a critical relative value'' for the theoretical study and numerical simulation of water vapor is actually equivalent to an infinite value for $\alpha$ in the assumed condensation probability function $k=(q/q_s)^{\alpha}$.

\vspace*{4mm}
{\bf \large \hspace{10mm} 3.2)  
\underline{Non-uniform 
saturated wet aerodynamic equations}
\vspace{2mm} \\ \hspace*{25mm}
\underline{assumed to be a 
function of condensation density.}
}
\vspace*{2mm}
\label{32_saturated_density}

According to the facts of cloud physics observation [5] and the above analysis, it is roughly concluded that the larger the concentration of hygroscopic condensed $\rho \, N$, the easier the water vapor will condense, and the lower is the relative humidity $(q/q_s)_0$ at which condensates can be found.

If you use $\sim$ to indicate a positive correlation, roughly
\begin{equation}
\nonumber
\hspace{0mm}
\frac{1}{\rho \, N}
\; \sim \;
(q/q_s)_0
\: ;
\hspace{20mm} (41) 
\end{equation}
In addition, from the expression of the condensation probability function $k=(q/q_s)^\alpha$ and the criterion (40), it can be seen that for the determined discovery condensation criterion $k_0$, there is a positive correlation between $\alpha$ and $(q/q_s)_0$ when condensation occurs, i.e.
\begin{equation}
\nonumber
\hspace{0mm}
\alpha
\; \sim \;
(q/q_s)_0
\: .
\hspace{20mm} (42) 
\end{equation}
Therefore, from (41) and (42):
\begin{equation}
\nonumber
\hspace{0mm}
\alpha
\; \sim \;
\frac{1}{\rho \, N}
\: .
\hspace{20mm} (43) 
\end{equation}
Based on the above analysis, and combined with the existing results of dynamic meteorology and atmospheric physics, we assume
\begin{equation}
\nonumber
\hspace{0mm}
\alpha
\; \sim \;
\frac{\beta}{\rho \, N}
\: ,
\hspace{20mm} (44) 
\end{equation}
\begin{equation}
\nonumber
\hspace{0mm}
\frac{dm}{dt}
\; \sim \;
0
\: ,
\hspace{20mm} (44) 
\end{equation}
\begin{equation}
\nonumber
\hspace{0mm}
\Lambda
\; \sim \;
\gamma \; \eta^{3/4} 
\: , \; \; \mbox{(from [6])}
\hspace{20mm} (46)
\end{equation}
where $\beta$ and $\gamma$ are the constant coefficients to be determined.
Then equations (15)-(25) can be reduced to non-uniform saturated wet aerodynamic equations, assuming that $\alpha$ is a function of the concentration of condensed nodules $\rho \: N$
\begin{equation}
\nonumber
\hspace{0mm}
\frac{d \, V}{dt}
\: + f \: K \times  V 
\; = \;
 - \: \frac{1}{\rho} \: \nabla P
\: ,
\hspace{20mm} (47) 
\end{equation}
\begin{equation}
\nonumber
\hspace{0mm}
\frac{d \, w}{dt}
\; = \;
 - \: \frac{1}{\rho} 
   \: \frac{\partial \, P}{\partial z}
\: - \: g
\: ,
\hspace{20mm} (48) 
\end{equation}
\begin{equation}
\nonumber
\hspace{0mm}
\frac{d \, T}{dt}
\: - \: \frac{1}{C_p} \: \frac{1}{\rho} 
   \: \frac{d \, P}{d t}
\; = \;
 - \: \frac{L}{C_p} \;
 \frac{
 d \left[ \:
 {(q/q_s)}^{\beta/(\rho \, N)} \: q_s
 \: \right]
 }{dt} 
\: ,
\hspace{20mm} (49) 
\end{equation}
\begin{equation}
\nonumber
\hspace{0mm}
P
\; = \;
 \rho \: R \: T
\: ,
\hspace{20mm} (50) 
\end{equation}
\begin{equation}
\nonumber
\hspace{0mm}
\frac{1}{\rho} \:
\frac{d \, \rho}{dt}
\: + \:
\nabla \, . \, V
\: + \:
\frac{\partial \, \omega}{\partial z}
\; = \;
 0
\: ,
\hspace{20mm} (51) 
\end{equation}
\begin{equation}
\nonumber
\hspace{0mm}
\frac{d \, q}{dt}
\; = \;
 \frac{
 d \left[ \:
 {(q/q_s)}^{\beta/(\rho \, N)} \; q_s
 \: \right]
 }{dt} 
\: ,
\hspace{20mm} (52) 
\end{equation}
\begin{equation}
\nonumber
\hspace{0mm}
- \:
 \frac{d q}{dt} 
\; = \;
\eta
\: ,
\hspace{20mm} (53) 
\end{equation}
\begin{equation}
\nonumber
\hspace{0mm}
\frac{d \, \rho \, N}{dt}
\; = \;
\nabla \: . \: 
\left(
\sigma \: . \: \nabla \, \rho \, N
\right)
\: - \: \gamma \: \rho \, N \; \eta^{3/4}
\: + \: V_g 
     \: \frac{\partial \, \rho \, N}{\partial z}
\: ,
\hspace{20mm} (54) 
\end{equation}
\begin{equation}
\nonumber
\hspace{0mm}
q_s
\; = \;
f_s  \left(P, T \right)
\: .
\hspace{20mm} (55) 
\end{equation}
Since equations (49), (52), and (54) relate to the condensation concentration of $\rho \: N$, equations (47)-(55) can discuss the role of condensation nuclei in large-scale motions involving water vapor.
In-depth research will be discussed separately.

It is also possible to make some new assumptions for $\alpha$ and $m$ according to different discussion objectives, which constitute a non-uniform saturated wet air dynamics equation.
I won't go into details here.

\vspace*{4mm}
{\bf \large 4) \underline{Conclusion and discussion}
}
\label{4_Conclusion}
\vspace*{2mm}

So far, this paper has abandoned the assumption that the condensation process starts at a certain critical relative value while introducing the condensation probability function, so that the wet air dynamics study can get rid of the unreasonable requirement that the whole atmosphere is always saturated. 
The non-uniform saturated wet air dynamics equations are given, and some simplified forms of the equations are given under certain assumptions.

These equations are similar and different from the traditional equations and have a more general form.
In fact, as long as $q=0$, $q=q_s$ or $\alpha=\infty$, these equations will degenerate into various traditional equations.

\vspace*{4mm}
{\bf \large \underline{References}}
\vspace*{2mm}
\label{References}

[1] Xie Yibing. 
The weather dynamics problem of wet baroclinic atmosphere. 
See: Rainstorm Collection. 
Changchun: Jilin People's Publishing House, 
1978. 1-15.

[2] Wang Liangming, Luo Huibang. 
Basic equations and main characteristics of saturated wet air dynamics. 
Acta Meteorologica Sinica, 
1980, 38 (1): 44-50.

[3] Wu Guoxiong, Cai Yaping, Tang Xiaojing. 
Development of wet vortices and inclined vortices. 
Acta Meteorologica Sinica, 
1995, 53 (4) : 387-405.

[4] Wang Xingrong, Shi Chunxi, Wang Zhongxing. 
Vertical coordinate transformation and wet 
aerodynamics equations under non-hydrostatic 
equilibrium conditions. 
Atmospheric Science, 
1997, 21 (5): 557-563.

[5] Meisen, B J. Cloud Physics (translated by the 
Institute of Atmospheric Physics, Chinese 
Academy of Sciences) Beijing: 
Science Press, 1978. 1-32.

[6] Li Zongkai, Pan Yunxian, Sun Runqiao. 
Principles and Applications of Air Pollution Meteorology. 
Beijing: Meteorological Press, 
1985. 380-394.

          \newpage 

\setcounter{equation}{0} 

\begin{center}
{\Large
{\bf Appendix~4:~}{\it A (crude and partial) translation  of the Chinese paper\/}\\
\vspace{2mm}
{\bf Wang} X. R.${}^{(1)}$ and M. {\bf Wei}${}^{(2)}$ (2007).\\
\vspace{1mm}
{\bf  \color{blue}
Theoretical analysis of non-uniform saturated moist potential
vorticity (NUSMPV) associated with the occurrence and development 
of weather systems.\\
}
\vspace{2mm}
{\bf
Journal of Tropical Meteorology, 2007, 23 (5), 459-566.
}
}
\end{center}
\vspace{0 mm}

(1) Anhui Institute of Meteorological Sciences, Anhui Hefei

(2) Sino-US Cooperation Remote Sensing Laboratory, 
   Nanjing University of Information Science and 
   Technology, Jiangsu Key Laboratory of Meteorological 
   Disasters, Nanjing, Jiangsu

\vspace{3 mm}

{\noindent \small WWW links:
\url{http://en.cnki.com.cn/Article_en/CJFDTotal-RDQX200705005.htm}
}

{\noindent \small
\url{http://www.itmm.org.cn/rdqxxb/ch/reader/view_abstract.aspx?file_no=20070506&flag=1}
}

{\noindent \small The PDF: $\:$
\url{http://www.itmm.org.cn/rdqxxb/ch/reader/create_pdf.aspx?file_no=20070506&year_id=2007&quarter_id=5&falg=1}
}

\vspace*{4mm}
{\bf \large \underline{Summary}}
\vspace*{2mm}
\label{Summary}

In terms of dynamic meteorological theory, whether large-scale, mesoscale or small-scale atmospheric motion, there is a fairly perfect discussion of the atmospheric motion evolution mechanism within the same scale.
However, discussions on the evolution mechanism of transitions between weather systems at different scales are very important, even because they are closely related to the emergence and development of weather systems, but are rarely involved.

In order to discuss this important transformation evolution mechanism, based on the derivation of the non-viscous non-uniform saturated wet vortex (NUSMPV) equation, firstly, the mathematical analysis of this equation proves: 
\begin{enumerate}
\item that in the non-uniform saturated wet adiabatic atmospheric motion, the dry NUSMPV of the unsaturated region between the region and the saturation region is not conserved, and 
\item that only in the subsaturated region ($0.78 < q/q_s<1$), discussion about the dimensionless form of NUSMPV shows that the atmospheric motion is divided into three Class, i.e. NUSMPV conservative, quasi-conservative and non-conservative motion.
\end{enumerate}

Then, the mathematical conditions are used to discuss the emergence conditions of various types of motion, the relationship between time and space, physical mechanism and some characteristics.

Finally, the conversion mechanism of the weather system between different scales is analyzed. It is pointed out that: the dynamic imbalance (A+B) between the vortex tube item (A) and the non-adiabatic heating item (B) related to the narrow NUSMPV($P_0$) is reduced, so that when the NUSMPV conservation condition is satisfied, atmospheric motion will continue to lose unbalanced energy mainly through a very fast dispersion adaptation process, from a small scale to a larger scale; and when $(A+B)/P_0$ increases, when the non-conservative condition of NUSMPV is satisfied, the atmospheric motion will mainly change from a larger scale to a smaller scale through a very fast exogenous excitation process.
Regardless of which of the above two conversion processes, once generated, it will continue until the atmospheric motion returns to the NUSMPV quasi-conservative motion state.

\vspace*{4mm}
{\bf \large 1)  \underline{Introduction}}
\vspace*{2mm}
\label{1_Intro}

At present, in terms of dynamic meteorological theory, whether large-scale, mesoscale or small-scale atmospheric motion, there is a fairly perfect discussion on the evolution mechanism of atmospheric motion within the same scale [1].
However, discussions on the evolution mechanism of transitions between weather systems of different scales are very important, even because they are closely related to the emergence and development of weather systems, but are rarely involved.

Ertel (1942) proposed the concept of vortex (DPV) and pointed out that DPV is strictly conserved in a non-viscous dry adiabatic airflow [2].
Hoskins et al. (1985) systematically analyzed the application of DPV in the study of atmospheric phenomena and introduced the concept of isentropic vortices [3].
DPV is not conserved when considering latent heat release.
However, by replacing the potential temperature $\theta$ with the equivalent potential temperature $\theta_e$, the wet vortex (MPV) is conserved during the non-viscous wet adiabatic saturation process.
Since Bennetts and Hoskins (1979) [4] and Emanuel (1979) [5] first proposed conditional symmetry instability (CSI) as a possible mechanism for the formation of frontal rainbands, the MPV concept has been widely applied to the baroclinic system CSI study.
Since then, people have relied on DPV conservation, reversibility (Robinson 1988) [6]; Davis (1992) [7]; Huo et al. (1999) [8] and the inaccessibility of vortex species (Haynes and Mcintyre (1990) [9], widely used to solve problems related to large-scale weather phenomena.

In recent years, on the basis of the conservation of the wet vortex (MPV) of the non-viscous wet adiabatic saturated airflow, Wu Guoxiong et al. proposed the theory of inclined vorticity development, pointing out that the vortex is more likely to occur on the steep wet isentropic surface [10,12].
Schubert et al. (2001) introduced the total density $\rho$ composed of dry air density and airborne water (water vapor and cloud), effective potential temperature 
$\theta_{\rho} = T_{\rho} \: (p_0/P)^{R_a/C_{pa}}$ 
(where effective temperature $T_{\rho} = p / (\rho \: R_a)$ and the well-known Ertel DPV's appropriate extension 
$P = \rho^{-1} (\: 2 \: \Omega + \nabla \times \theta_{\rho} \: )$, 
extending the MPV conservation and reversibility principle  under balanced airflow conditions by Ooyama's (1990, 2001) [13-14] in non-hydrostatic precipitation mode [15]; and according to the immobility of MPV substances, Gao et al. (2002) pointed out that MPV anomalies are an indicator for tracking storm systems [16].

In the large-scale DPV or MPV study, many scholars (Hoskins and Berridford (1988) [17]; Keyser and Rotunno (1990) [18]; Cao and Cho (1995) [19]; Cho and Cao (1998) [ 20]; Gao et al. (2002) [16]) noted the generation and abnormality of DPV and MPV [16-20].
In numerical simulations, DPV inversion techniques, including DPV anomaly processing, have been widely used to improve the initial conditions of the model (Reed et al (1992)) [21].

It is precisely because of the importance of DPV and MPV in the study of large-scale phenomena that many researchers attempt to study mesoscale phenomena using DPV or MPV anomalies (Fritsch and Maddox (1981) [22] ; Fritsch et al. 1994 [23] ; Davis and Weisman (1994) [24] ; Skamarock et al. (1994) [25] ; Gray et al. (1998) [26]).
Although there are some studies on the abnormalities in DPV or MPV (Raymond and Jiang (1990) [27]; Raymond (1992) [28]; Shutts and Gray (1994) [29]; Fulton et al. (1995) [30]), However, these studies, especially the mechanisms by which MPV anomalies are generated, are still not very clear, and there is little discussion about the relationship between DPV or MPV anomalies and weather system conversions at different scales.

As early as 1965, Ye Yongzheng pointed out: ``The atmosphere is generally quasi-adapted, that is, quasi-equilibrium, but the imbalance of imbalance in the atmosphere is often present, it is the driving force for generating atmospheric motion.''
And ``when the weather system occurs strongly, it is often accompanied by a very strong imbalance (i.e., not adapting) phenomenon.''
Therefore, it is considered that ``the study of the atmospheric imbalance process and its formation mechanism is also an important issue'' [31].
In 1984, Wang Xingrong proposed that the (obvious) geostrophic deviation rapid **excitation** and rapid **adaptation** are the causes of atmospheric circulation adjustment, and the (obvious) geostrophic deviation generation process is the PV non-conservation process [32].
Since then, Wang Xingrong has made a series of studies on issues related to PV and MPV non-conservation processes [33-38].

Especially in recent years, it is noted that during the weather, the atmosphere is often in a state of non-uniform saturation, that is, some places are saturated, some places are not saturated, and the previous wet air mode generally assumes that the condensation process starts at a certain critical value. 
The assumptions force a discontinuous $\delta$ function in the kinetic equations, making the dynamic analysis of the boundary between the condensation zone and the non-condensation zone extremely difficult.
In response to this defect, Wang Xingrong et al. (1995 [39], 1997 [40], 1999 [41]) replaced the $\delta(0,1)$ function by introducing a continuous condensation probability function $(q/q_s)^k$.
On this basis, Wang Xingrong et al. (1998) [33] derived the non-uniform saturated wet vortex (NUSMPV) equation.
Through the dynamic analysis of sudden severe weather, Wang Xingrong et al. (1998) [34] pointed out that sudden severe weather has a fierce and rapid {\bf excitation process} and a relatively stable and slow {\bf development process}.
Subsequently, Wang Xingrong et al. (1999) [35] pointed out that the mid-latitude ground rotation deviation {\bf excitation process} is a NUSMPV non-conservation process caused by the dynamic imbalance of various factors under the condition that NUSMPV tends to zero.

Based on the above research, this paper will explore the mechanism of weather system conversion at different scales.

\vspace*{4mm}
{\bf \large 2)  \underline{Basic equations}}
\vspace*{2mm}
\label{2_Basic_equ}

In the non-viscous non-uniform saturated atmospheric motion, according to the literature [39-41], by introducing a continuous condensation probability function $(q/q_s)^k$ instead of the $\delta(0,1)$ function, a non-uniform saturation thermodynamics equation can be obtained
\begin{equation}
\nonumber
\hspace{40mm} 
C_p \: \frac{T}{\theta} \: 
\frac{d \theta}{dt} 
\; = \;
- \: L \: \frac{d}{dt}
\left[ \:
{\left(
\frac{q}{q_s}
\right)}^k
\: q_s \:
\right]
\; + \; Q_d
\: .
\hspace{30mm} (1) 
\end{equation}
If you define generalized wet temperature $\theta^{\ast}$
\begin{equation}
\nonumber
\hspace{60mm} 
\theta^{\ast} 
\; = \;
\theta \:
\exp\left[\:
\frac{L}{C_p}\:
\frac{q_s}{T}\:
{\left(
\frac{q}{q_s}
\right)}^k
\: \right]
\: ,
\hspace{40mm} (2) 
\end{equation}
the formula (1) can be expressed as
\begin{equation}
\nonumber
\hspace{60mm}
\frac{d \theta^{\ast}}{dt} 
\; = \;
\frac{\theta^{\ast}}{C_p \: T} \: Q_d
\; = \;
Q
\: ,
\hspace{50mm} (3) 
\end{equation}
where $Q$ is defined as the generalized non-adiabatic heating, including the latent heat release.

Using the similar derivation used by Cao and Cho (1995) [19] and Wu Guoxiong et al. [10], the momentum equation for the Euler form writes
\begin{equation}
\nonumber
\hspace{40mm} 
\frac{\partial \vec{V}}{\partial t} 
\: + \:
\nabla
\: \left(
\frac{V^2}{2} \: + \: \phi
\right)
\: - \:
\vec{V} \: \times \vec{\zeta}_a 
\; = \;
- \, \alpha \: \nabla p
\hspace{30mm} (4) 
\end{equation}
and the vorticity equation
\begin{equation}
\nonumber
\hspace{40mm} 
\frac{\partial}{\partial t} \: \vec{\zeta}_a
\: - \:
\nabla \: \times \: 
\: \left(
\vec{V} \: \times \: \vec{\zeta}_a
\right)
\; = \;
 \nabla p \: \times \: \nabla \alpha
\hspace{30mm} (5) 
\end{equation}
where $\vec{V}$ is the 3D wind vector and $\phi$ is the potential function of the external force (gravity), and 
$\vec{\zeta}_a = \nabla \times \vec{V} + 2 \: \vec{\Omega}$ 
is the three-dimensional vorticity vector.

The use of $\nabla \: \theta^{\ast}$ to multiply (5), the use of equation (3), and the use of the vector relations
\begin{equation}
\nonumber
\hspace{30mm} 
\nabla \theta^{\ast} \, . \: 
\nabla \times
\left(
\vec{V} \times \vec{\zeta}_a
\right)
\; = \;
- \:
\nabla \, . \: 
\left[ \: 
\nabla \,\theta^{\ast} 
\times
\left(
\vec{V} \times \vec{\zeta}_a
\right)
\: \right]
\hspace{25mm} (6) 
\end{equation}
and
\begin{equation}
\nonumber
\hspace{30mm} 
\nabla \theta^{\ast} \, \times \:
\left(
\vec{V} \times \vec{\zeta}_a
\right)
\; = \;
\vec{V} \:
\left(
\vec{\zeta}_a \: . \: \nabla \theta^{\ast}
\right)
\: + \:
\vec{\zeta}_a \:
\left(
\frac{\partial \theta^{\ast}}{\partial t} 
\: - \: Q
\right)
\hspace{20mm} (7) 
\end{equation}
lead to
\begin{equation}
\nonumber
\hspace{10mm} 
\left( 
  \frac{\partial}{\partial t} 
  \: + \: 
  \vec{V} \: . \: \nabla
\right)
\left(
 \vec{\zeta}_a \: . \: \nabla \theta^{\ast}
\right)
   \: + \:
\left(
 \vec{\zeta}_a \: . \: \nabla \theta^{\ast}
\right)
 \nabla  \: . \: \vec{V}
\; = \;
\left(
\nabla p \: \times \: \nabla \alpha
\right)
\: . \: \nabla \theta^{\ast}
\; + \;
\vec{\zeta}_a \: . \: \nabla  Q
\; \: .
\hspace{10mm} (8) 
\end{equation}

It is then possible to multiply the two sides of the equation (8) with the specific volume $\alpha$ and to use the continuous equation
\begin{equation}
\nonumber
\frac{d \, \alpha}{d t} 
\: - \: 
\alpha \: \nabla  \: . \: \vec{V}
\; = \;
0 \; ,
\end{equation}
leading to the NUSMPV equation
\begin{equation}
\nonumber
\hspace{40mm} 
{\color{black}
\boxed{ \:
\frac{d \, P_m}{d t} 
\; = \;
\alpha \:
\left( 
\nabla p \: \times \: \nabla \alpha
\right)
\: . \: \nabla \theta^{\ast}
\; + \;
\alpha \; \:
\vec{\zeta}_a \: . \: \nabla  Q
\: }
} \; ,
\hspace{30mm}
{\color{black}
 (9) 
}
\end{equation}
where
\begin{equation}
\nonumber
\hspace{40mm} 
{\color{black}
\boxed{ \:
P_m
\; = \;
\alpha \: \: \vec{\zeta}_a \: . \nabla \theta^{\ast}
\: }
}
\; ,
\hspace{30mm} 
{\color{black}
(9a) 
}
\end{equation}
It should be noted that the NUSMPV equation is consistent with the general MPV equation expression, but the NUSMPV equation calculates the generalized wet temperature $\theta^{\ast}$ instead of the equivalent temperature $\theta_e$. 
Equation (9) also has the following form:
\begin{equation}
\nonumber
\hspace{40mm} 
{\color{black}
\boxed{ \:
\frac{d \, P_m}{d t} 
\; = \; P_m \;
\frac{( A \: + \: B )}{P_0} 
\: }
} \; ,
\hspace{30mm}
{\color{black}
 (10) 
}
\end{equation}
where
\begin{equation}
\nonumber
\hspace{20mm} 
{\color{black}
\boxed{ \:
A
\; = \; 
 \left\{
 \frac{R_d}{p} 
 {\left(
 \frac{p}{p_0} 
 \right)}^{-R_d/C_p}
 \right\}
\:
\left(
\nabla p \: \times \: \nabla \theta
\right)
\: . \: \nabla \theta^{\ast}
\: }
} \; ,
\hspace{15mm}
{\color{black}
 (10a) 
}
\end{equation}
\begin{equation}
\nonumber
\hspace{50mm} 
{\color{black}
\boxed{ \:
B
\; = \; 
\vec{\zeta}_a \: . \nabla Q
\: }
} \; ,
\hspace{40mm}
{\color{black}
 (10b) 
}
\end{equation}
\begin{equation}
\nonumber
\hspace{50mm} 
{\color{black}
\boxed{ \:
P_0
\; = \; 
\vec{\zeta}_a \: . \nabla \theta^{\ast}
\: }
} \; .
\hspace{40mm}
{\color{black}
 (10c) 
}
\end{equation}
Note that, in the derivation of Eq.(10), the equation of state and 
$\alpha = (R_d/p) \: \theta \: (p/p_0)^{-R_d/C_p}$ are used. 
In equation (10), 
$A$ is the force tube term related to the baroclinic and generalized wet temperature gradient; 
$B$ is the non-adiabatic term, 
$P_0$ is the product of the three-dimensional vorticity vector 
$\vec{\zeta}_a$ 
($\vec{\zeta}_a = \nabla \times \vec{V} + 2 \: \vec{\Omega}$)
and the generalized wet temperature gradient vector 
$\nabla \: \theta^{\ast}$, 
which we call the narrow NUSMPV.

For a non-viscous wet adiabatic atmospheric motion ($\nabla \: Q = 0$, $B = 0$), and for either an absolutely dry zone ($q = 0$, $(q /q_s)^k = 0$, $\theta^{\ast} = \theta$) or a fully saturated wet zone ($q=q_s$, $(q/q_s)^k =1$, $\theta^{\ast} = \theta_{se}$), then $A$ is zero, and so NUSMPV is strictly conserved.
However, in the general unsaturated zone, the transition zone between the absolute dry zone and the saturated wet zone, $A$ is not zero, i.e. NUSMPV is not conserved.

In the (just) sub-saturated region (namely where $0.78<q/q_s<1.0$), $\theta^{\ast}$ depends on the humidity $((q/q_s)^k \neq 1$) and NUSMPV is not conserved.
In other unsaturated regions (namely where $q/q_s<0.78$), NUSMPV is almost nearly conserved because $k$ is large enough to cause $(q/q_s)^k \approx 0$.

Therefore, the force tube effect ($A$) of baroclinic and generalized wet temperature gradients can play a significant role in the generation of NUSMPV only in the sub-saturated region.
This is consistent with the results of the numerical simulations of Cho and Cao (1998) [20].

\vspace*{4mm}
{\bf \large 3) \underline{Three 
types of atmospheric motion with}
\vspace{2mm} \\ \hspace*{9mm}
\underline{different 
conservation characteristics}
}
\vspace*{2mm}
\label{3_Atmospheric_motion}
\vspace*{2mm}

There are three types of atmospheric motion according to different conservation characteristics: when $(A+B)/P_0$ is small enough to be negligible, NUSMPV can be considered to be conserved; when $(A+B)/P_0$ has a moderate value, NUSMPV can be considered as Quasi-conservative; when $(A+B)/P_0$ is large enough to dominate, NUSMPV can be considered as non-conservative.
The dimensionless equation of equation (10) can be expressed as
\begin{equation}
\nonumber
\hspace{0mm} 
               {\color{black}
               \boxed{ \:
\varepsilon \:
\frac{\partial \, P_{m1}}{\partial t_1} 
\: + \:
R \: 
\left[ \:
u_1 \: \frac{\partial \, P_{m1}}{\partial x_1}
 + 
v_1 \: \frac{\partial \, P_{m1}}{\partial y_1}
\: \right]
\: + \:
\frac{W}{f \: H} \: 
\left(
w_1 \: \frac{\partial \, P_{m1}}{\partial z}
\right)
\; = \; 
P_{m1} \:
\left(
f^{-1} \: 
\frac{( A \: + \: B )}{P_0} 
\right)
                \: }
                }
\; .
\hspace{3mm}
{\color{black}
 (11) 
}
\end{equation}
{\color{black}
where $\varepsilon$ is the Kibil (??) number and $R$ is the Rossby number.
}

In order to identify whether NUSMPV is conserved, the following criteria can be used:
if
\begin{equation}
\nonumber
\hspace{0mm} 
               {\color{black}
               \boxed{ \:
O(\varepsilon) 
\; = \; 
\max
\left(
O(R), 
O\left(\frac{W}{f \: H}  \right)
\right)
\; > \;
O\left(
  f^{-1} \: 
  \frac{( A \: + \: B )}{P_0} 
\right)
                \: }
                }
\; ,
\hspace{3mm}
{\color{black}
 (12a) 
}
\end{equation}
then NUSMPV is conserved; 
and if
\begin{equation}
\nonumber
\hspace{0mm} 
               {\color{black}
               \boxed{ \:
O(\varepsilon) 
\; = \; 
\max
\left(
O(R), 
O\left(\frac{W}{f \: H}  \right)
\right)
\; = \;
O\left(
  f^{-1} \: 
  \frac{( A \: + \: B )}{P_0} 
\right)
                \: }
                }
\; ,
\hspace{3mm}
{\color{black}
 (12b) 
}
\end{equation}
then NUSMPV is quasi-conservative;
and if
\begin{equation}
\nonumber
\hspace{0mm} 
               {\color{black}
               \boxed{ \:
O(\varepsilon) 
\; = \; 
O\left(
  f^{-1} \: 
  \frac{( A \: + \: B )}{P_0} 
\right)
\; > \;
\max
\left(
O(R), 
O\left(\frac{W}{f \: H}  \right)
\right)
                \: }
                }
\; ,
\hspace{3mm}
{\color{black}
 (12c) 
}
\end{equation}
then NUSMPV is not conserved.

\vspace*{4mm}
{\bf \large \hspace{10mm} 3.1)  \underline{
NUSMPV conservation movement.
}}
\vspace*{2mm}
\label{31_Conservation_movement}

Most observational facts and dimensional analysis show that the spatial scale of atmospheric motion generally has a corresponding relationship with the degree of damage between the pressure gradient force, Corioli force, inertial force and gravity. The smaller the scale, the greater the damage.

Therefore, any NUSMPV-conserved atmospheric motion with the spatial scale 
$\max[O(R),$ $ O(W/fH)] =O(10^n)$ satisfying equation (12a) can be regarded as having a larger spatial scale 
$\max[O(R ), O(W/fH)] =O(10^{n-1})$ 
a perturbation of atmospheric motion.
There is a fairly mature discussion about this movement (Ye Yizheng and Li Maicun (1965) [31]).
This is a very fast adaptation process in which the imbalance between the pressure gradient force, the Corioli force, the inertial force and the gravity disappears due to the dispersion of the gravity wave and the sound wave.
Through this adaptation process, plus the friction effect, the smaller scale 
$(\max[O(R), O(W/fH)] =O(10^n))$ 
atmospheric motion will be converted to a larger scale 
$(\max[O(R) , O(W/fH)] = O(10^{n-1}))$ atmospheric motion.
Once this process is produced, it will continue until the atmospheric motion no longer satisfies equation (12a).
In fact,  if atmospheric motion is converted from the smaller scale $(\max[O(R), O(W/fH)] =O(10^n))$ to a larger scale $(\max[O(R), O(W/fH)] =O(10^{n-1}))$, then atmospheric motion is very fast.

\vspace*{4mm}
{\bf \large \hspace{10mm} 3.2)  \underline{
NUSMPV non-conservation movement.
}}
\vspace*{2mm}
\label{32_Non_Conservation_movement}

Because the NUSMPV non-conservative atmospheric motion of any spatial scale 
$\max[O(R),$ $ O(W/fH)] =O(10^n)$ satisfies the equation (12c),
\begin{equation}
\nonumber
\hspace{0mm} 
               {\color{black}
               \boxed{ \:
O\left(
  f^{-1} \: 
  \frac{( A \: + \: B )}{P_0} 
\right)
\; > \;
\max
\left(
O(R), 
O\left(\frac{W}{f \: H}  \right)
\right)
                \: }
                }
\; ,
\hspace{3mm}
{\color{black}
 (13a) 
}
\end{equation}
\begin{equation}
\nonumber
\hspace{0mm} 
               {\color{black}
               \boxed{ \:
O(\varepsilon) 
\; > \;
\max
\left(
O(R), 
O\left(\frac{W}{f \: H}  \right)
\right)
                \: }
                }
\; ,
\hspace{3mm}
{\color{black}
 (13b) 
}
\end{equation}
\begin{equation}
\nonumber
\hspace{0mm} 
               {\color{black}
               \boxed{ \:
O(\varepsilon) 
\; = \; 
O\left(
  f^{-1} \: 
  \frac{( A \: + \: B )}{P_0} 
\right)
                \: }
                }
\; .
\hspace{3mm}
{\color{black}
 (13c) 
}
\end{equation}
Equation (13a) is a necessary and sufficient condition for the occurrence of NUSMPV unconserved atmospheric motion.
It can be seen from equation (10c) that $P_0$ mainly depends on the factor 
$\nabla \theta^{\ast}$ and the factor $\vec{\zeta}_a$.
In actual atmospheric motion, no matter how large or how small $A$ and $B$ are, as long as the characteristic value $O(P_0)$ of $P_0$ is small enough to satisfy
\begin{equation}
\nonumber
\hspace{0mm} 
               {\color{black}
               \boxed{ \:
O(P_0)
 \; < \;
O\left(
  f^{-1} \: 
  ( A \: + \: B )
\right)
\; / \;
\max
\left(
O(R), 
O\left(\frac{W}{f \: H}  \right)
\right)
                \: }
                }
\; .
\hspace{3mm}
{\color{black}
 (14) 
}
\end{equation}
Then equation (13a) will be satisfied, and NUSMPV will appear when the air is not conserved.
In other words, if intense non-adiabatic heating associated with lightning and thunderstorms is not considered, then when $P_0$ is sufficiently close to zero, the disturbance caused by $(A+B)$ can excite NUSMPV unconserved atmospheric motion.
It is in accordance with this condition that, through the dynamic analysis of sudden severe weather, Wang Xingrong et al. (1999 [35]) pointed out that the sudden disaster weather excitation process is caused by various factors under the condition that NUSMPV tends to zero. 
The NUSMPV non-conservative process is caused by dynamic imbalance.
According to this condition, Wang Xingrong et al. also carried out other applications (2001 [36], 2002 [37], 2003 [38]).

For a non-conserved atmospheric motion with a spatial scale of NUSMPV, equation (13b) shows that its time scale is much smaller than the general atmospheric motion with the same spatial scale, which is a fast process.

Equation (13c) indicates that the time scale of NUSMPV non-conservative atmospheric motion depends mainly on the eigenvalues $O(P_0)$ and $(A+B)$ of $P_0$.

When NUSMPV does not conserve atmospheric motion, equation (13a) is satisfied. Therefore equations (9) and (10) can be simplified to
\begin{equation}
\nonumber
\hspace{10mm} 
               {\color{black}
               \boxed{ \:
\frac{\partial \, P_m}{\partial t} 
\; = \;
\alpha \:
\left( 
\nabla p \: \times \: \nabla \alpha
\right)
\: . \: \nabla \theta^{\ast}
\; + \;
\alpha \; \:
\vec{\zeta}_a \: . \: \nabla  Q
\; = \;
P_m \: \frac{( A \: + \: B )}{P_0}
              \: }
               }
\; .
\hspace{20mm}
{\color{black}
 (15) 
}
\end{equation}
Using $P_m$ and $P_0$ given by the expressions (9a) and (10c), equation (15) becomes
\begin{equation}
\nonumber
\hspace{10mm} 
               {\color{black}
               \boxed{ \:
\frac{P_0}{\alpha} \;
\frac{\partial \, \alpha}{\partial t} 
\: + \:
\frac{\partial \, \vec{\zeta}_a}{\partial t} 
\: . \: \nabla \theta^{\ast}
\: + \:
\vec{\zeta}_a
\: . \: \frac{\partial \, \nabla \theta^{\ast}}{\partial t} 
\; = \;
 A \: + \: B
              \: }
               }
\; .
\hspace{20mm}
{\color{black}
 (16) 
}
\end{equation}
If you do not consider the intense non-adiabatic heating associated with lightning and thunderstorms, then
\begin{equation}
\nonumber
\hspace{10mm} 
               {\color{black}
               \boxed{ \:
O\left(
\frac{\Delta \alpha}{\alpha}
\right)
\; < \;
 10^0
              \: }
               }
\; .
\hspace{20mm}
{\color{black}
 (17) 
}
\end{equation}
Using equations (14) and (17) leads to
\begin{equation}
\nonumber
\hspace{10mm} 
               {\color{black}
               \boxed{ \:
\frac{\partial \, P_0}{\partial t} 
\; = \;
\frac{\partial \, \vec{\zeta}_a}{\partial t} 
\: . \: \nabla \theta^{\ast}
\: + \:
\vec{\zeta}_a
\: . \: \frac{\partial \, \nabla \theta^{\ast}}{\partial t} 
\; = \;
 A \: + \: B
              \: }
               }
\; .
\hspace{20mm}
{\color{black}
 (18) 
}
\end{equation}
Since the necessary and sufficient condition for the non-conservative atmospheric motion of NUSMPV is that the eigenvalue $O(P_0)$ of $P_0$ is sufficiently small, equation (18) indicates that the non-conserved atmospheric motion of NUSMPV is a sufficiently small eigenvalue $O(P_0)$ that $P_0$ suddenly becomes larger. 
Process (not because it was moved from the upstream).
In this process $(A+B)$, the vorticity field, and $\nabla \theta^{\ast}$ are changed, and thus $P_0$ is changed.
As a result, it caused a disturbance with a smaller scale 
$(\max[O (R),O(W/fH)]=O(10^{n+m}))$ than the original scale 
$(\max[O(R),O(W/fH)]=O(10^n))$.

Once this process is produced, it will continue until the atmospheric motion no longer satisfies equation (13a) and satisfies the following relationship
\begin{equation}
\nonumber
\hspace{0mm} 
               {\color{black}
               \boxed{ \:
O\left(
  f^{-1} \: 
  \frac{( A \: + \: B )}{P_0} 
\right)
\; \leq \;
\max
\left(
O(R), 
O\left(\frac{W}{f \: H}  \right)
\right)
                \: }
                }
\hspace{3mm}
{\color{black}
 (19) 
}
\end{equation}
That is, the process stops until NUSMPV becomes conserved or quasi-conserved.
Through this process, a large-scale atmospheric movement 
$(O(\varepsilon)=O(f^{-1} (A+B)/P_0 ) > \max[O(R),$ $O(W/fH)]=O(10^n))$ 
is converted into a smaller-scale atmospheric movement 
$(O(\varepsilon)=\max[O(R), O(W/fH)]=O(10^{n+m}) 
\geq O(f^{-1} (A+B)/P_0 ) > O(10^n)))$.
We call this process an 
{\color{black} \bf exciting process}.

It should be pointed out that during the excitation process, the {\color{black} \bf adaptation process} will also occur at the same time, but it can be ignored.

In addition, the dynamic and physical characteristics of NUSMPV unconserved atmospheric motion at different latitudes and different scale systems are different.

In 1998, Wang Xingrong et al. [33] used the non-uniform saturated wet air dynamics equations to discuss the dynamic and physical characteristics of the non-conserved atmospheric motion of NUSMPV in the background of large-scale mid-latitude systems. 
It is pointed out that the large-scale NUSMPV at mid-latitude is not during the conservation process.
The vertical non-adiabatic heating and latent heat vertical change the stratification, which changes the NUSMPV, which changes the divergence field and causes the ground rotation deviation, thus stimulating mesoscale and small-scale atmospheric motion.
NUSMPV non-conservative atmospheric motion in the context of other scale systems are for further exploration in the future.

\vspace*{4mm}
{\bf \large \hspace{10mm} 3.3)  \underline{
NUSMPV quasi-conservation movement.
}}
\vspace*{2mm}
\label{33_Quasi_Conservation_movement}

Because the NUSMPV quasi-conservative atmospheric motion at any spatial scale $(\max[O(R),$ $O(W/fH)]=O(10^n))$ satisfies the equation (12b),
\begin{equation}
\nonumber
\hspace{0mm} 
               {\color{black}
               \boxed{ \:
O(\varepsilon) 
\; = \;
\max
\left(
O(R), 
O\left(\frac{W}{f \: H}  \right)
\right)
                \: }
                }
\; ,
\hspace{3mm}
{\color{black}
 (20a) 
}
\end{equation}
\begin{equation}
\nonumber
\hspace{0mm} 
               {\color{black}
               \boxed{ \:
\max
\left(
O(R), 
O\left(\frac{W}{f \: H}  \right)
\right)
\; = \;
O\left(
  f^{-1} \: 
  \frac{( A \: + \: B )}{P_0} 
\right)
                \: }
                }
\; ,
\hspace{3mm}
{\color{black}
 (20b) 
}
\end{equation}
\begin{equation}
\nonumber
\hspace{0mm} 
               {\color{black}
               \boxed{ \:
O(\varepsilon) 
\; = \; 
O\left(
  f^{-1} \: 
  \frac{( A \: + \: B )}{P_0} 
\right)
                \: }
                }
\; .
\hspace{3mm}
{\color{black}
 (20c) 
}
\end{equation}
Equation (20) shows that the NUSMPV quasi-conservative atmospheric motion is not only related to the time scale and spatial scale, but also depends on $(A+B)/P_0$.
It can be seen from equation (20b) that the change in $(A+B)/P_0$ is very slow. 
Otherwise, the equation (20b) cannot be satisfied and the NUSMPV conservation atmospheric motion or the NUSMPV non-conservative atmospheric motion occurs.
This means that the NUSMPV quasi-conservative atmospheric motion time scale and spatial scale do not change or change very slowly. 
Therefore, this is the evolution of the quasi-conservation of NUSMPV caused by $(A+B)/P_0$.
This is a slow process.
In this process, not only $(A+B)/P_0$ can not be ignored, but also the dispersion and nonlinear advection are equally important.

Therefore, there are three main processes affecting the NUSMPV quasi-conservative atmospheric motion: (1) the excitation process caused by $(A+B)$; (2) the disturbance caused by the excitation process disappears by the dispersion of gravity waves and acoustic energy. Adaptation process (additional processes such as friction and turbulence are added to the actual atmospheric motion); (3) Nonlinear advection process.

\vspace*{4mm}
{\bf \large 4) \underline{Conversion 
mechanism and application of }
\vspace{2mm} \\ \hspace*{9mm}
\underline{weather system between 
different scales}
}
\vspace*{2mm}
\label{4_Conversion_mechanism}

From the above analysis, it can be seen that the NUSMPV conservation atmospheric motion and the NUSMPV non-conservative atmospheric motion are fast processes, and the NUSMPV quasi-conservative atmospheric motion is a slow process.
Therefore, most of the observed NUSMPV of atmospheric motion are quasi-conservative.
Therefore, it can be assumed that the initial atmospheric motion is the NUSMPV quasi-conservative atmospheric motion, that is, the $(A+B)/P_0$ associated with the NUSMPV change corresponds to the scale of the initial atmospheric motion [see equation (12b)]

If a physical phenomenon, such as a cut in a water vapor channel or a sudden change in a non-adiabatic heating field, changes $(A+B)/P_0$ so that $(A+B)/P_0$ no longer corresponds to the scale of atmospheric motion, That is to say, formula (12b) is no longer satisfied, and there is
\begin{equation}
\nonumber
\hspace{0mm} 
               {\color{black}
               \boxed{ \:
O(\varepsilon) 
\; = \;
\max
\left(
O(R), 
O\left(\frac{W}{f \: H}  \right)
\right)
\; > \;
O\left(
  f^{-1} \: 
  \frac{( A \: + \: B )}{P_0} 
\right)
                \: }
                }
\; ,
\hspace{3mm}
{\color{black}
 (21) 
}
\end{equation}
or
\begin{equation}
\nonumber
\hspace{0mm} 
               {\color{black}
               \boxed{ \:
O(\varepsilon) 
\; = \;
O\left(
  f^{-1} \: 
  \frac{( A \: + \: B )}{P_0} 
\right)
\; > \;
\max
\left(
O(R), 
O\left(\frac{W}{f \: H}  \right)
\right)
                \: }
                }
\; .
\hspace{3mm}
{\color{black}
 (22) 
}
\end{equation}
Then, according to the above discussion, the scale of atmospheric motion will be converted

From Section 3.1 on the discussion of NUSMPV's conservation of atmospheric motion, it can be known that when $(A+B)/P_0$ is reduced, so that the NUSMPV conservation atmospheric motion of any scale of equation (21) is present, a very fast adaptation process will occur. Through the energy dispersion of gravity waves and sound waves, a small-scale $(\max[O(R),O(W/fH)]=O(10^n))$ atmospheric motion will become a larger-scale $(max[O(R),O(W/fH)]=O(10^{n-1}))$ atmospheric motion, and this process will continue until the equation (12b) is satisfied, that is, the atmospheric motion returns to a status of quasi-conservation.

From the discussion of Section 3.2 on the non-conserved atmospheric motion of NUSMPV, we can know: when $(A+B)/P_0$ increases, so that any scale of NUSMPV that satisfies equation (22) does not conserve atmospheric motion, a very fast excitation process It will appear that the $(A+B)$ forcing changes the vorticity field and $\nabla \theta^{\ast}$ changes the narrow sense NUSMPV, causing a perturbation with a smaller scale 
$(\max[O(R) ,O(W/fH)]=O(10^{n+m}))$ 
than the original scale 
$(\max[O(R)O(W/fH)]=O(10^n))$, 
i.e. a larger scale 
$(\max[O(R), O(W/fH)]=O(10^n))$ 
atmospheric motion will it becomes a smaller scale 
$(\max[$ $O(R),O(W/fH)]>O(10^n))$ 
atmospheric motion, and this process will continue until the equation (12b) is satisfied, that is, the atmospheric motion returns to the quasi-conservative state.

According to the above analysis, in the early stage of the sudden disaster weather ({\bf exciting phase}), the atmospheric motion often shifts from a large scale to a medium-to-small scale within a certain height range. Therefore, the conditions for the NUSMPV non-conservative atmospheric motion must be satisfied.
Therefore, the conditions for the occurrence of some sudden-onset disasters and the indicators of the near-forward predictions can be achieved by Wang Xingrong et al. (43), which has been tried and achieved with some success [43]) by dynamic analysis of the conditions of the non-conservation process of the wet vortex (NUSMPV).

For example, through the discussion of the conditions for the occurrence of the non-conservation process of the wet vortex (NUSMPV), Wang Xingrong et al. (2006) [43] theoretically proved that: before the mid-latitude sudden rainstorm occurs,
(1) there must be in the deep convection height range, 
$\partial \theta^{\ast} / \partial z$
 rapidly approaches zero; 
(2) the baroclinic vector has a component along the generalized wet temperature gradient, 
(3) the nearly saturated unsaturated layer is thick enough to achieve deep convection 
(4) located near clouds and clouds containing large amounts of water vapor and liquid water. 
Hefei twice in 1999 (22:00-24:00 on June 15 and rainfall of 68.5 mm; 14:00-15:00 on August 1 and 80.5 mm of rainfall). 
Sudden heavy rains do reflect the above four necessary two common characteristics of the condition; 
(5) Before the rainstorm occurred, there was a deep double-layer dry area over Hefei (a two-layer continuous no-data area in the deep convective height range on the Doppler radar VWP map).
First, as the wet and cold air at the top of the upper dry zone begins to slide, the upper dry zone begins to disappear from top to bottom. 
Then, when the sliding air stops sliding and begins to rise, the lower dry zone begins to disappear, and finally the second layer is dried. When the area disappears completely, the mesoscale convective system that causes sudden heavy rain suddenly appears.
This is a possible aura characteristic that satisfies the first necessary condition of a sudden heavy rain and makes the deep convective atmosphere nearly saturated (but not guaranteed to be unsaturated).
(6) At the same time, on the cloud map, the south-long cold front cloud belt splits into two parts in the vicinity of Hefei, so that there is almost no cloud and fog at the break, and there are two clouds containing a large amount of liquid water and water vapor. After the cold front cloud passes through Hefei, the mesoscale convective system that brings sudden heavy rain breaks out at the break with almost no clouds and fog.
This is another possible aura characteristic that satisfies the second and fourth necessary conditions of a sudden torrential rain and places the deep convective atmosphere at an unsaturated state.
The two possible aura characteristics are combined to meet the third necessary condition of a sudden rainstorm.

\vspace*{4mm}
{\bf \large 5)  \underline{
Conclusion and discussion}}
\vspace*{2mm}
\label{5_Conclusion_discussion}

Based on the accurate derivation of the non-uniform saturated wet vortex (NUSMPV) equation, firstly, according to this equation, in the frictionless wet adiabatic atmospheric motion, the NUSMPV of the unsaturated zone between the dry zone and the saturated zone is not conserved. And further proves that NUSMPV is not conserved only in the subsaturated region ($0.78<q/qs<1$), while in other unsaturated regions, NUSMPV may be nearly conserved.

Then, according to the principle of conservation of conservation, through the NUSMPV dimensionless form discussion, the atmospheric motion is divided into three categories, namely NUSMPV {\bf conservative}, {\bf quasi-conservative} and {\bf non-conservative} motion.

Then we discuss the emergence of various types of sports, the relationship between time and space, physical mechanisms and some characteristics.

Finally, the conversion mechanism of weather system between different scales is analyzed. It is pointed out that the dynamic imbalance $(A+B)/P_0$ between the vortex tube item (A) and the non-adiabatic heating item (B) related to the narrow NUSMPV($P_0$) $P_0$ is reduced, so that when the NUSMPV conservation condition is satisfied, atmospheric motion will continue to lose unbalanced energy mainly through very fast gravity wave and acoustic wave dispersion adaptation process, from small scale to larger scale, and when $(A+B)/P_0$ increases, so that the NUSMPV non-conservation condition is satisfied, the atmospheric motion will change the exogenous excitation process of NUSMPV mainly by changing the vorticity and $\nabla \theta^{\ast}$ very quickly by $(A+B)$, and conversion from larger scale to smaller scale.

Regardless of the two conversion processes described above, once generated, they will continue until the atmospheric motion returns to the NUSMPV quasi-conservative motion state.

It should be pointed out that the above discussion is preliminary and basic. In order to reveal the physical mechanisms of various types of atmospheric motion at different scales and to describe their dynamic and physical characteristics, there are still many questions to answer.

As far as the current situation is concerned, there are some mature discussions about the NUSMPV conservation atmospheric motion and adaptation process (the same as the MPV conservation atmospheric motion and adaptation process).
For the NUSMPV quasi-conservative atmospheric motion and evolution process, although some scholars many fruitful studies have been carried out at various angles, but because the NUSMPV quasi-conserved atmospheric motion and evolution processes are related to many nonlinear factors, many problems are still quite vague and require further research.
For NUSMPV non-conservative atmospheric motion and the concept of the {\bf excitation} process and the {\bf conversion} mechanism of the weather system between different scales has only been proposed in recent years. The work we have done is mainly to discuss them theoretically (although some application attempts have been made), so it is almost blank. It is yet to be drawn in the future to attract more interested scholars to conduct comprehensive and extensive research.

\vspace*{4mm}
{\bf \large \underline{References}}
\vspace*{2mm}
\label{References}

[1] Ye Yizheng, Li Chongyin. 
Dynamic Meteorology[M]. 
Beijing: China Science Press, 1998: 1-126

[2] ERTEL H. 
Ein neuer hydrodynamischer wirbelsatz[J]. 
Meteorolog Zeitchr Braunschweig. 
1942, 59: 271-281.

[3] HOSKINS B J, MCINTYRE M E, ROBERTSON R W. 
On the use and significance of isentropic potential 
vorticity amps[J]. 
Quarterly Journal of the Royal Meteorological Society. 
1985, 111: 877-946.

[4] BENNETTS D A, HOSKINS B J. 
Conditional symmetric instability a possible 
explanation for frontal rainbands[J]. 
Quarterly Journal of the Royal Meteorological Society,
1979, 105: 945-962.

[5] EMANUEL K A. 
Inertial instability and mesoscale convective systems. 
Part I: Linear theory of inertial instability in rotating 
viscous fluids[J].
Journal of the Atmospheric Sciences, 
1979, 36: 2425-2449.

[6] ROBINSON W A. 
Analysis of LIMS data by potential vorticity inversion[J]. 
Journal of the Atmospheric Sciences, 
1988, 45: 2319-2342.

[7] DAVIS C A. 
A potential vorticity diagnosis of the importance 
of initial structure and condensational heating in 
observed extratropical cyclogenesis[J]. 
Monthly Weather Review, 
1992, 120: 2409-2428.

[8] HUO Z, ZHANG D L, GYAKUM J R. 
Interaction of potential vorticity anomalies in 
extratropical cyclogenesis, part I: Static 
piecewise inversion[J]. 
Monthly Weather Review, 
1999, 127: 2546-2561.

[9] HAYNES P H, MCINTYRE M E. 
On the conservation and impermeability theorems for 
potential vorticity[J]. 
Journal of the Atmospheric Sciences, 
1990, 47: 2021-2031.

[10] Wu Guoxiong, Cai Yaping, Tang Xiaotong. 
Development of wet vorticity and inclined vorticity[J]. 
Journal of Meteorology, 
1995, 53 (4): 387-405.

[11] Wu Guoxiong, Liu Huazhu. 
Full-scale vertical vorticity tendency equation and 
development of inclined vorticity[J]. 
Acta Meteorologica Sinica, 
1999, 57(1): 1-15.

[12] Wu Guoxiong, Liu Wei. 
Thermal adaptation, overcurrent and subtropical high 
I. Thermal adaptation and overcurrent [J].
Atmospheric Sciences, 
2000, 24(4): 433-446.

[13] OOYAMA K V. 
A thermodynamic foundation for modeling the moist 
atmosphere[J]. 
Journal of the Atmospheric Sciences. 
1990, 47: 2580-2593.

[14] OOYAMA K V. 
A dynamic and thermodynamic foundation for modeling 
the moist atmosphere with parameterized microphysics[J]. 
Journal of the Atmospheric Sciences, 
2001, 58: 2073-2102.

[15] SCHUBERT W H, SCOTT A H, MATTHEW GARCIA, et al. 
Potential Vorticity in a Moist Atmosphere[J]. 
Journal of the Atmospheric Sciences, 
2001, 58: 3148-3157.

[16] GAO Shouting, LEI Ting, ZHOU Yushu. 
Moist potential vorticity anomaly with heat and 
mass forcings in torrential rain system[J]. 
Chinese Physics Letters, 
2002, 19: 878-880.

[17] HOSKINS B J, BERRIDFORD P. 
A potential-vorticity perspective of the storm of 
15-16 October 1987[J]. 
Weather, 1988, 43: 122-129.

[18] KEYSER D, ROTUNNO R. 
On the formation of potential-vorticity anomalies 
in upper-level jet-front systems[J]. 
Monthly Weather Review,
1990, 118: 1914-1921.

[19] CAO Zuohao, CHO Han-ru. 
Generation of moist potential vorticity in extratropical 
cyclones[J]. 
Journal of the Atmospheric Sciences, 
1995, 52(18): 3263-3281.

[20] CHO Han-ru, CAO Zuohao. 
Generation of moist vorticity in extratropical cyclones. 
Part II: Sensitivity to moisture distribution[J]. 
Journal of the Atmospheric Sciences, 
1998, 55: 595-610.

[21] REED R J, STOELINGA M T, KUO Y H. 
A model aided study of the origin and evolution of the 
anomalously high potential vorticity in the inner region 
of a rapidly deepening marine cyclone[J]. 
Monthly Weather Review, 
1992, 120: 893-913.

[22] FRITSCH J M, MADDOX R A. 
Convectively drive mesoscale weather system aloft. 
Part I: Observations[J]. 
Journal of Applied Meteorology,
1981, 20(1): 9-19.

[23] FRITSCH J M, MURPHYJ D, KAIN J S. 
Warm core vortex amplification over land[J]. 
Journal of the Atmospheric Sciences, 
1994, 51: 1780-1807.

[24] DAVIS C A, WEISMAN M L. 
Balanced dynamics of mesoscale vortices in simulated 
convective systems[J]. 
Journal of the Atmospheric Sciences, 
1994, 51: 2005-2030.

[25] SKAMAROCK W C, WEISMAN M L, KLEMP J B. 
Three-dimensional evolution of simulatede long-lived 
squall lines[J]. 
Journal of the Atmospheric Sciences, 
1994, 51: 2563-2584.

[26] GRAY M E B, SHUTTS G J, CRAIG G C. 
The role of mass transfer in describing the dynamics 
of mesoscale convective system[J]. 
Quarterly Journal of the Royal Meteorological Society, 
1998, 124: 1183-1207.

[27] RAYMOND D J, JIANG H. 
A theory for long-lived mesoscale convective systems[J]. 
Journal of the Atmospheric Sciences, 
1990, 47: 3067-3077.

[28] Raymond D J. 
Nonlinear balance and potential-vorticity thinking at 
large Rossby number[J]. 
Quarterly Journal of the Royal Meteorological Society, 
1992, 118: 987-1 015.

[29] SHUTTS G. J, GRAY M E B. 
A numerical modelling study of the geostrophic adjustment 
process following deep convection[J]. 
Quarterly Journal of the Royal Meteorological Society, 
1994, 120: 1145-1178.

[30] FULTON S R, SCHUBERT W H, HAUSMAN S A. 
Dynamical adjustment of mesoscale convective anvils[J]. 
Monthly Weather Review, 1995,
123: 3215-3226.

[31] Ye Yizheng, Li Maicun. 
Adaptation problems in atmospheric motion [M]. 
Beijing: Science Press, 1965: 1-126.

[32] Wang Xingrong. 
Dynamic Analysis of Circulation Adjustment Mechanism[J]. 
Plateau Meteorology, 1984, 3 (1): 27-35.

[33] Wang Xingrong, Wang Zhongxing, Shi Chunqi. 
On the non-conservation of wet vortices in atmospheric 
motion[J]. 
Meteorological Science, 1998, 13(2): 136-141.

[34] Wang Xingrong. 
Exploring sudden disasters from the trigger mechanism 
of unstable energy[J]. 
Journal of Natural Disasters, 1998, 7(1): 11-15.

[35] Wang Xingrong, Wu Kejun, Chen Xiaoping, et al. 
Dynamics analysis of the characteristics and occurrence 
conditions of sudden disasters[J]. 
Journal of Nanjing Institute of Meteorology, 
1999, 22(4): 711-715.

[36] Wang Xingrong, Shang Yu, Yao Yeqing, et al. 
The connection characteristics and dynamics between 
the subtropical high activity and the lunar month[J]. 
Journal of Tropical Meteorology, 
2001, 17(4): 423-428.

[37] Wang Xingrong. 
From the "wet vorticity is not conserved" to study 
sudden severe weather, new ways and new methods for 
predicting catastrophic natural disasters [M] 
Xiangshan Science Conference 133th Academic Symposium. 
Beijing: China Science and Technology Press, 
2002: 128-131.

[38] WANG Xingrong,YAO Yeqing, YU Shang, et al. 
Analyses of errors in medium-term numerical forecast 
products for subtropical high 1998[J]. 
Journal of Tropical Meteorology (English version), 
2003, 9(1): 105-112.

[39] Wang Xingrong, Wu Kejun. 
Discussion on Some Problems of Wet Air Dynamics[J]. 
Meteorological Science, 
1995, 15(1): 9-17.

[40] Wang Xingrong, Shi Chunxi, Wang Zhongxing. 
Vertical coordinate transformation and wet air dynamics 
equations under non-hydrostatic equilibrium conditions[J]. 
Journal of Atmospheric Sciences, 
1997, 21(5): 557-563.

[41] Wang Xingrong, Wu Kejun, et al. 
Introduction of condensation probability function 
and non-uniform saturated wet air dynamics equations[J]. 
Journal of Tropical Meteorology, 
1999, 15(1): 64-70.

[42] MASON B J. 
The physics of clouds[M]. 
Oxford: Oxford University Press, 
1971: 1-671.

[43] Wang Xingrong, Zheng Yuanyuan, Gao Shouting, et al. 
Possible aura characteristics of sudden torrential 
rain in mid-latitudes[J]. 
Journal of Tropical Meteorology, 
2006, 22(6): 612-617.

 \end{document}